\documentclass[iop,apj,numberedappendix,twocolumn, tighten]{aastex62}
\usepackage{natbib}
\usepackage{graphicx}
\usepackage{amsmath,amsfonts}
\usepackage{multirow}
\usepackage{xspace}
\usepackage[normalem]{ulem}
\usepackage{booktabs}
\usepackage{colortbl}
\usepackage{hyperref}
\usepackage[nameinlink]{cleveref}
\usepackage[encapsulated]{CJK}
\usepackage{ucs}
\usepackage{float}
\usepackage[utf8x]{inputenc}
\newcommand{\cntext}[1]{\begin{CJK}{UTF8}{gbsn}#1\end{CJK}}
\usepackage{eht}
\usepackage{newtxtext} 

\usepackage{lineno}

\graphicspath{{./figures}}

\newcommand{\uv}{$(u,v)$\xspace}

\newcommand{\greyrule}{\arrayrulecolor{black!30}\midrule\arrayrulecolor{black}}

\setcitestyle{notesep={; }}

\def\sgra{Sgr~A$^{\ast}$\xspace}

\def\lsim{\mathrel{\raise.3ex\h box{$<$\kern-.75em\lower1ex\hbox{$\sim$}}}}
\def\gsim{\mathrel{\raise.3ex\hbox{$>$\kern-.75em\lower1ex\hbox{$\sim$}}}}
\def\gtwid{\mathrel{\raise.3ex\hbox{$>$\kern-.75em\lower1ex\hbox{$\sim$}}}}
\def\proptwid{\mathrel{\raise.3ex\hbox{$\propto$\kern-.75em\lower1ex\hbox{$\sim$}}}}

\begin{document}
\title{\textbf{Selective dynamical imaging of interferometric data }} 
\shorttitle{Selective dynamical imaging of interferometric data}
\author[0000-0003-4914-5625]{Joseph Farah}
\affiliation{Las Cumbres Observatory, 6740 Cortona Drive, Suite 102, Goleta, 
CA 93117-5575, USA}
\affiliation{Department of Physics, University of California, Santa Barbara, 
CA 93106-9530, USA}

\author[0000-0002-6429-3872]{Peter Galison}
\affiliation{Black Hole Initiative at Harvard University, 20 Garden Street, Cambridge, MA 02138, USA}
\affiliation{Department of History of Science, Harvard University, Cambridge, MA 02138, USA}
\affiliation{Department of Physics, Harvard University, Cambridge, MA 02138, USA}

\author[0000-0002-9475-4254]{Kazunori Akiyama}
\affiliation{Massachusetts Institute of Technology Haystack Observatory, 99 Millstone Road, Westford, MA 01886, USA}
\affiliation{National Astronomical Observatory of Japan, 2-21-1 Osawa, Mitaka, Tokyo 181-8588, Japan}
\affiliation{Black Hole Initiative at Harvard University, 20 Garden Street, Cambridge, MA 02138, USA}

\author[0000-0003-0077-4367]{Katherine L. Bouman}
\affiliation{Black Hole Initiative at Harvard University, 20 Garden Street, Cambridge, MA 02138, USA}
\affiliation{Center for Astrophysics | Harvard \& Smithsonian, 60 Garden Street, Cambridge, MA 02138, USA}
\affiliation{California Institute of Technology, 1200 East California Boulevard, Pasadena, CA 91125, USA}

\author[0000-0003-4056-9982]{Geoffrey C. Bower}
\affiliation{Institute of Astronomy and Astrophysics, Academia Sinica, 645 N. A'ohoku Place, Hilo, HI 96720, USA}

\author[0000-0003-2966-6220]{Andrew Chael}
\affiliation{Princeton Center for Theoretical Science, Jadwin Hall, Princeton University, Princeton, NJ 08544, USA}
\affiliation{NASA Hubble Fellowship Program, Einstein Fellow}

\author[0000-0002-8773-4933]{Antonio Fuentes}
\affiliation{Instituto de Astrof\'{\i}sica de Andaluc\'{\i}a-CSIC, Glorieta de la Astronom\'{\i}a s/n, E-18008 Granada, Spain}

\author[0000-0003-4190-7613]{Jos\'e L. G\'omez}
\affiliation{Instituto de Astrof\'{\i}sica de Andaluc\'{\i}a-C\'{\i}SIC, Glorieta de la Astronom\'{\i}a s/n, E-18008 Granada, Spain}

\author[0000-0003-4058-9000]{Mareki Honma}
\affiliation{Mizusawa VLBI Observatory, National Astronomical Observatory of Japan, 2-12 Hoshigaoka, Mizusawa, Oshu, Iwate 023-0861, Japan}
\affiliation{Department of Astronomical Science, The Graduate University for Advanced Studies (SOKENDAI), 2-21-1 Osawa, Mitaka, Tokyo 181-8588, Japan}
\affiliation{Department of Astronomy, Graduate School of Science, The University of Tokyo, 7-3-1 Hongo, Bunkyo-ku, Tokyo 113-0033, Japan}

\author[0000-0002-4120-3029]{Michael D. Johnson}
\affiliation{Black Hole Initiative at Harvard University, 20 Garden Street, Cambridge, MA 02138, USA}
\affiliation{Center for Astrophysics | Harvard \& Smithsonian, 60 Garden Street, Cambridge, MA 02138, USA}

\author{Yutaro Kofuji}
\affiliation{Mizusawa VLBI Observatory, National Astronomical Observatory of Japan, 2-12 Hoshigaoka, Mizusawa, Oshu, Iwate 023-0861, Japan}
\affiliation{Department of Astronomy, Graduate School of Science, The University of Tokyo, 7-3-1 Hongo, Bunkyo-ku, Tokyo 113-0033, Japan}

\author[0000-0002-2367-1080]{Daniel P. Marrone}
\affiliation{Steward Observatory and Department of Astronomy, University of Arizona, 933 N. Cherry Ave., Tucson, AZ 85721, USA}

\author[0000-0003-1364-3761]{Kotaro Moriyama}
\affiliation{Massachusetts Institute of Technology Haystack Observatory, 99 Millstone Road, Westford, MA 01886, USA}
\affiliation{Mizusawa VLBI Observatory, National Astronomical Observatory of Japan, 2-12 Hoshigaoka, Mizusawa, Oshu, Iwate 023-0861, Japan}

\author[0000-0002-1919-2730]{Ramesh Narayan}
\affiliation{Black Hole Initiative at Harvard University, 20 Garden Street, Cambridge, MA 02138, USA}
\affiliation{Center for Astrophysics | Harvard \& Smithsonian, 60 Garden Street, Cambridge, MA 02138, USA}

\author[0000-0002-5278-9221]{Dominic W. Pesce}
\affiliation{Black Hole Initiative at Harvard University, 20 Garden Street, Cambridge, MA 02138, USA}
\affiliation{Center for Astrophysics | Harvard \& Smithsonian, 60 Garden Street, Cambridge, MA 02138, USA}

\author[0000-0003-3826-5648]{Paul Tiede}
\affiliation{Department of Physics and Astronomy, University of Waterloo, 200 University Avenue West, Waterloo, ON, N2L 3G1, Canada}
\affiliation{Waterloo Centre for Astrophysics, University of Waterloo, Waterloo, ON, N2L 3G1, Canada}

\author[0000-0002-8635-4242]{Maciek Wielgus}
\affiliation{Max-Planck-Institut f\"ur Radioastronomie, Auf dem H\"ugel 69, D-53121 Bonn, Germany}

\author[0000-0002-4417-1659]{Guang-Yao Zhao}
\affiliation{Instituto de Astrof\'{\i}sica de Andaluc\'{\i}a-CSIC, Glorieta de la Astronom\'{\i}a s/n, E-18008 Granada, Spain}

\shortauthors{Farah et al.}

\correspondingauthor{$^\dag$Joseph R. Farah}

\email{josephfarah@ucsb.edu} 

\collaboration{The Event Horizon Telescope Collaboration}


\author[0000-0002-9371-1033]{Antxon Alberdi}
\affiliation{Instituto de Astrof\'{\i}sica de Andaluc\'{\i}a-CSIC, 
Glorieta de la Astronom\'{\i}a s/n, E-18008 Granada, Spain}

\author{Walter Alef}
\affiliation{Max-Planck-Institut f\"ur Radioastronomie, Auf dem H\"ugel 69, D-53121 Bonn, Germany}

\author[0000-0001-6993-1696]{Juan Carlos Algaba}
\affiliation{Department of Physics, Faculty of Science, Universiti Malaya, 50603 Kuala Lumpur, Malaysia}

\author[0000-0003-3457-7660]{Richard Anantua}
\affiliation{Black Hole Initiative at Harvard University, 20 Garden Street, Cambridge, MA 02138, USA}
\affiliation{Center for Astrophysics $|$ Harvard \& Smithsonian, 60 Garden Street, Cambridge, MA 02138, USA}
\affiliation{Department of Physics \& Astronomy, The University of Texas at San Antonio,
One UTSA Circle, San Antonio, TX 78249, USA}

\author[0000-0001-6988-8763]{Keiichi Asada}
\affiliation{Institute of Astronomy and Astrophysics, Academia Sinica, 11F of 
Astronomy-Mathematics Building, AS/NTU No. 1, Sec. 4, Roosevelt Rd, Taipei 10617, Taiwan, R.O.C.}

\author[0000-0002-2200-5393]{Rebecca Azulay}
\affiliation{Departament d'Astronomia i Astrof\'{\i}sica, Universitat de Val\`encia, C. Dr. Moliner 50, E-46100 Burjassot, Val\`encia, Spain}
\affiliation{Observatori Astronòmic, Universitat de Val\`encia, C. Catedr\'atico Jos\'e Beltr\'an 2, E-46980 Paterna, Val\`encia, Spain}
\affiliation{Max-Planck-Institut f\"ur Radioastronomie, Auf dem H\"ugel 69, D-53121 Bonn, Germany}

\author[0000-0002-7722-8412]{Uwe Bach}
\affiliation{Max-Planck-Institut f\"ur Radioastronomie, Auf dem H\"ugel 69, D-53121 Bonn, Germany}

\author[0000-0003-3090-3975]{Anne-Kathrin Baczko}
\affiliation{Max-Planck-Institut f\"ur Radioastronomie, Auf dem H\"ugel 69, D-53121 Bonn, Germany}

\author{David Ball}
\affiliation{Steward Observatory and Department of Astronomy, University of Arizona, 933 N. Cherry Ave., Tucson, AZ 85721, USA}

\author[0000-0003-0476-6647]{Mislav Balokovi\'c}
\affiliation{Yale Center for Astronomy \& Astrophysics, Yale University, 52 Hillhouse Avenue, 
New Haven, CT 06511, USA} 

% ...

\begin{abstract}
Recent developments in very-long-baseline interferometry (VLBI) have made it possible for the Event Horizon Telescope (EHT) to resolve the innermost accretion flows of the largest supermassive black holes on the sky. 
The sparse nature of the EHT's \uv coverage presents a challenge when attempting to resolve highly time-variable sources. We demonstrate that the changing \uv coverage of the EHT can contain regions of time over the course of a single observation that facilitate dynamical imaging. These optimal time regions typically have projected baseline distributions that are approximately angularly isotropic and radially homogeneous. We derive a metric of coverage quality based on baseline isotropy and density that is capable of ranking array configurations by their ability to produce accurate dynamical reconstructions. 
We compare this metric to existing metrics in the literature, and investigate their utility by performing dynamical reconstructions on synthetic data from simulated EHT observations of sources with simple orbital variability.
We then use these results to make recommendations for imaging the 2017 EHT \sgra dataset. 
\end{abstract}


\section{Introduction}
\label{sec:introduction}
Interferometric astronomical observations offer much larger resolving power than do single telescopes, with the interferometric resolution depending on the distance between the elements rather than the diameters of the individual apertures. Since an interferometer probes the Fourier transform of an on-sky source (and not the source image itself), the placement, selection and availability of baselines to maximize coverage of the \uv plane is an important and open optimization problem inherent to the interferometric image synthesis. For very-long-baseline interferometry (VLBI), which represents some of the most extreme interferometric observations, the arrays are generally very sparse, and optimized placement or selection of baselines becomes critical to recovering reliable information about the source.

For any ground-based interferometric observations,
including VLBI, Earth rotation causes each individual baseline to trace an elliptical path in the \uv plane. Such ``Earth rotation aperture synthesis'' can be used to improve the \uv coverage of sparse arrays. Standard VLBI imaging uses this additional coverage under the assumption that the target source structure remains unchanged for the duration of the observation. For the cases of sources that vary on shorter timescales (e.g., the Galactic Center, x-ray binaries, microquasars), ``snapshot images'' can be produced using only time intervals over which the source can be considered static \citep[e.g.,][]{Miller2019,Massi2012,Marti-Vidal2011}. In the most extreme scenarios, only instantaneous coverage may be suitable for use in static image reconstructions, severely limiting the \uv coverage and resulting image quality. Such a case is the main concern of this paper. Moreover, this limitation will generally result in portions of an observation that are better than others for imaging because the snapshot coverage is time-dependent.

In particular, directionally-biased \uv coverage results in a non-isotropic resolution that complicates interpretation of the reconstructed source geometry. The \uv coverage of a given array varies with the declination of the source, due to its dependence on the projected baseline lengths and orientations. The impact of source declination on the \uv coverage varies throughout a night of observation as the target rises and sets. A network of baselines which produces an evenly distributed \uv coverage of an equatorial source will be be East-West biased when targeting a northern source. Observations where the source morphology and orientation of the source are unknown cannot exploit \textit{a priori} knowledge to determine whether a directionally-biased \uv configuration will be able to reproduce source structure reliably in an image. Therefore, for cases where the source morphology is unknown, the optimal \uv coverage for producing high-quality image reconstructions will be those that are approximately isotropic in angular distribution.  

Periods of sustained isotropic \uv coverage may allow multiple high-quality image reconstructions on short timescales to be strung together, resulting in ``dynamical'' reconstructions. For instance, the most basic dynamical reconstruction can be produced by joining a series of snapshot images. The quality of such a reconstruction will be time-dependent due to the evolving snapshot coverage, and it may be necessary to exclude periods with poor \uv coverage. In periods of time where the \uv coverage is optimal, the coverage used to produce each snapshot of a dynamical reconstruction will be sufficiently isotropic to reproduce source features on snapshot timescales, allowing for the production of reconstructions that adequately recover the evolution of highly time-variable sources. By contrast, periods of sub-optimal \uv coverage may be unable to provide high-quality reconstructions on snapshot timescales, making the algorithmic detection and subsequent flagging of these periods important for the production of meaningful dynamical reconstructions.

The challenges of observing a variable source with a sparse interferometer can be exacerbated by poor coverage geometry. In particular, given the shortage of available facilities supporting high frequency radioastronomical observations, corresponding VLBI arrays exhibit substantiantial variance in coverage quality in time. 

The Event Horizon Telescope (EHT) is a unique VLBI network of telescopes that exploits the full diameter of the Earth and the performance at the challenging 1.3 mm (230 GHz) wavelength to achieve the required angular resolution for horizon-scale tests of general relativity for the largest black holes on the sky (see e.g., \citealt{Johannsen2010, Psaltis2015, Psaltis2018}). The EHT operates in millimeter wavelengths, the optimal range which enables the resolution of the \sgra black hole shadow and reduces the impact of the interstellar medium scattering effects dominant for longer wavelengths, while still being observable and manageable in the radio interferometric framework. Operating at 230 GHz, the EHT achieved a resolution of ${\sim}$20-25 $\mu$as, which the EHT Collaboration (EHTC) used to produce the first images of a supermassive black hole (\citealt{EHT1,EHT2,EHT3,EHT4,EHT5,EHT6}). These data and the resulting images were used to estimate a mass of $M\approx6.5\times10^9 M_\odot$ for the supermassive black hole in M87 (\citealt{EHT6}), based on an observed angular shadow diameter of ${\sim}42\mu$as (\citealt{EHT1, EHT4, EHT6}).

During the 2017 campaign, the EHT also observed the radio source \sgra in the Galactic Center (\citealt{PaperI, PaperII, PaperIII, PaperIV, PaperV, PaperVI}), associated with a supermassive black hole with $M\approx4.1\times10^6 M_\odot$ (\citealt{gravity_2018_s2,Do2019, PaperVI}). The expected mass-to-distance  ratio  ($M/D$) of  \sgra yields a predicted angular shadow diameter of ${\sim}$50 $\mu$as \citep{gravity_2018_hotspot}, and a minimum variability timescale (light crossing time) of $GM/c^3 \approx 20$ seconds. The corresponding timescale for M87 is ${\sim}$1600 times longer due to the larger mass. Indeed, structural variability of the M87 shadow has been reported on timescales from $\sim$1 week \citep{EHT4} to several years \citep{Wielgus2019}. The rapid minimum variability timescale of \sgra combined with the extreme sparsity of the EHT presents an urgent and unique need to to characterize the effects of time-dependent instantaneous \uv coverage.

In this paper, we develop a procedure for selective imaging of highly variable sources. In \autoref{sec:model_data} and \autoref{sec:imaging}, we summarize the synthetic data generation and imaging methods used herein. In \autoref{sec:demo}, we show the limitations of imaging in sparse and uneven coverage. In \autoref{sec:alternatives}, we survey several metrics capable of ranking \uv coverage quality. Additionally, we derive a novel isotropy-based metric which addresses the limitations described in \autoref{sec:demo}. In \autoref{sec:application_2017}, we apply these metrics to the 2017 EHT coverage of \sgra, validate their ability to predict reconstruction quality from \uv coverage geometry, and make recommendations for selective dynamical imaging of the 2017 EHT \sgra dataset. In \autoref{sec:inter-day}, we briefly discuss the utility of coverage metrics in ranking and selecting between different available observing periods. Finally, in \autoref{sec:conclusions}, we summarize our results.

\section{Model definition and synthetic data generation}
\label{sec:model_data}
In order to test the ability of various EHT array configurations to recover source variability in different observation periods, we designed and generated synthetic data for three different models. The models are chosen due to their structural similarity (in image and visibility domain) to expected images of \sgra. Similarity between the model and data was characterized as either displaying time variability or producing a Bessel-function Fourier representation with nulls between 2 and 4 G$\lambda$ and between $6$ and $9$ G$\lambda$. The models are described in \autoref{sec:models}. The synthetic data generation is expanded upon in \autoref{sec:toy_model_synth}.

\subsection{Models}
Here, we describe the suite of models used to test the effects of coverage on reconstructions. Examples of each model with CLEAN beam convolution can be seen in the first row of \autoref{fig:toy_models}.

\label{sec:models}
\subsubsection{Rotating elliptical Gaussian}
\label{sec:toy_model_gauss}

The rotating elliptical Gaussian model is generated using a bivariate exponential with major axis full-width-half-maximum (FWHM) $\Gamma_a$, minor axis FWHM $\Gamma_b$, and overall flux density $A$:
\begin{equation}
    I(x, y) =
      \frac{4A\ln2}{\pi  \Gamma_a\Gamma_b }
      \exp{\left(
        -4\ln2\left[
          \left(\frac{x}{\Gamma_a}\right)^2 +
          \left(\frac{y}{\Gamma_b}\right)^2 
        \right]
      \right)}.
\end{equation}
The image is padded and rotated by polar angle $\varphi$ after the model is generated. By default, the overall flux density $A$ was set to 1.0 Jy. Periods for the rotation ranged between 30 and 1000 minutes (longer than one night of observation).

\subsubsection{Ring and orbiting hotspot}
\label{sec:toy_model_ring}

The ring model is generated by the subtraction of two concentric uniform-brightness disks, equivalent to the crescent model described in \citet{Kamruddin2013}, with the parameters $a=b=0$. The positive disk has a radius of 25 $\mu$as and the subtracted disk has a radius of  18 $\mu$as. All ring models in this paper use these parameters with the exception of the ring model in \autoref{fig:vuln_radial}, which has a subtracted disk radius of 20 $\mu$as. The ring model is used with two synthetic data tests: the static ring (with no hotspot) and the dynamic ring (a static ring plus an orbiting hotspot, referred to as \texttt{ring+hs}). The underlying ring has a diameter of 50 $\mu$as and a flux density of 1.0 Jy. A hotspot total flux density of 0.25~Jy and a FWHM of 10 $\mu$as is added to the image, centered on the ring. After construction, the total flux density of each image is normalized to 1.0 Jy. In the dynamic ring model, the orbiting hotspot is centered on the ring and circularly orbiting at a distance of 21.5 $\mu$as with periods of 30 and 270 minutes. The static ring model is a special case of the ring+hotspot model with the flux density of the hotspot set to zero.

\begin{figure}[h]
\centering
\includegraphics[width=\columnwidth]{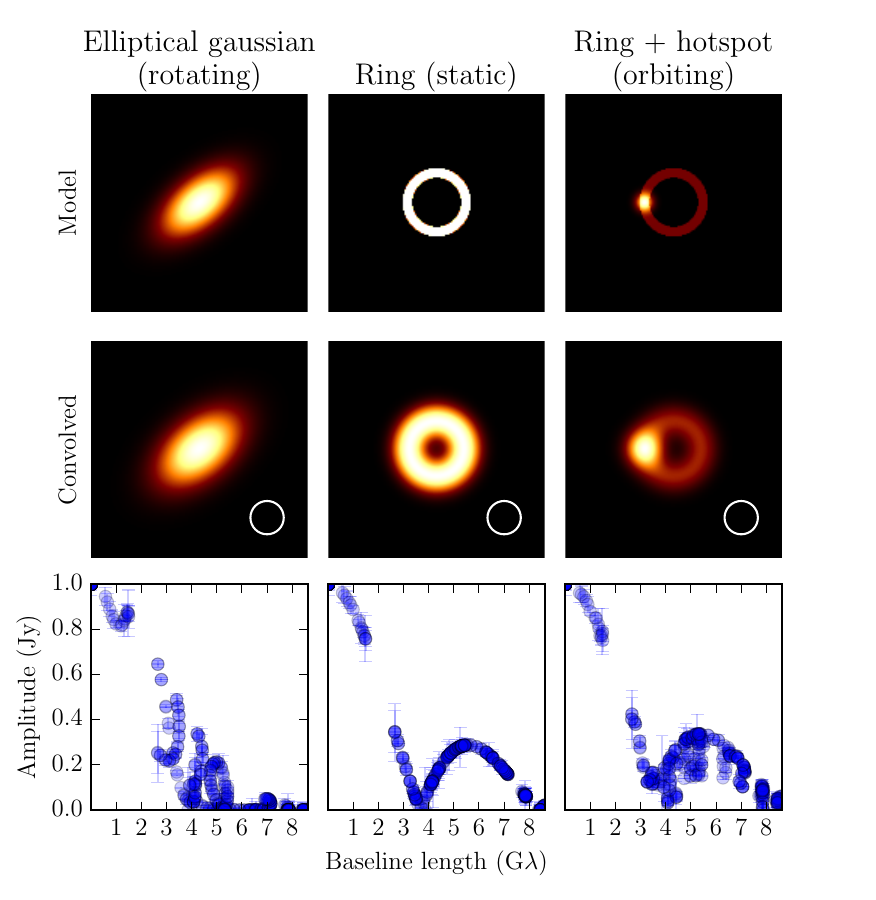}
\caption{The three synthetic models detailed in \autoref{sec:model_data} are displayed. The first row shows each model as seen on-sky at 6 UT, just after the observation begins. The second row shows the model convolved with an 18 $\mu$as diameter CLEAN beam. The white circle in the lower right shows the size of the beam. The third row shows the measured visibility amplitudes as a function of baseline length for the entire observation. A static imaging routine would fit to the full set of these amplitudes; however, a dynamical imaging routine only attempts to fit to small chunks of the full dataset at any one time.} 
\label{fig:toy_models}
\end{figure}

\subsection{Synthetic data generation}
\label{sec:toy_model_synth}

Synthetic data were generated based on April 7 of the 2017 EHT coverage using the \texttt{eht-imaging} library \citep{Chael2018}. A simulated observation corresponded to approximately 11.5 hours of observing on all available baselines. Snapshot images from the simulated sources and the resulting amplitudes and closure phases can be seen in \autoref{fig:toy_models} and \autoref{sec:sample_data}.  

Observation parameters (e.g., right ascension and declination of the source, observing frequency, bandwidth) were duplicated from the 2017 EHT survey of \sgra. The stations included in the simulated observations were the Atacama Large Millimeter Array (ALMA), the Atacama Pathfinder Experiment (APEX), the Large Millimeter Telescope (LMT), the James Clerk Maxwell Telescope (JCMT), the Submillimeter Array (SMA), the IRAM 30-meter telescope on Pico Veleta (PV), and the South Pole Telescope (SPT). 
Weather complications were ignored and the simulated observations assumed all stations were observing the source for the chosen night. 

All data were generated with thermal noise only. Realistic values for the thermal noise power were based on estimates from the real 2017 EHT data. No other noise, scattering, leakage, or simulated gain errors were applied to the simulated visibilities.

\section{Imaging approaches}
\label{sec:imaging}
Since the interferometric measurements are often incomplete in the Fourier domain, the inverse problem of reconstructing an image from the observed data set is usually under-determined. 
Consequently, the image reconstruction requires prior information, assumptions, or constraints to derive a reasonable image from the infinite number of possibilities that can explain the measurements.

The two most popular categories of imaging methodologies are inverse modeling (e.g., \texttt{CLEAN}) and forward modeling (e.g., regularized maximum likelihood). See \citet{EHT4} for a general overview of the two methods. For time-variable sources, both approaches may allow for more effective reconstructions of dynamic structures than snapshot imaging by including assumptions or constraints on temporal variations of the source structure in addition to the spatial properties regularized in static imaging \citep[e.g.][]{EHT4}.
One such way imposes a temporal similarity constraint between images at different times, and between each time snapshot and the time-averaged structure.
In the following Subsections, we briefly describe each dynamical reconstruction method used in this paper. See \cite{PaperIII} for more details.

\subsection{Inverse-modeling approaches}
\label{sec:clean}

Imaging of radio interferometric data is traditionally carried out through \texttt{CLEAN} deconvolution algorithms \cite[e.g.,][]{Hogbom_1974,Clark1980}. These inverse modeling approaches iteratively deconvolve the effects associated with the limited sampling of the \uv plane, corresponding to the interferometer's point source response (the so-called ``dirty beam'') to the inverse Fourier transform of the measured visibilities, commonly referred to as the ``dirty image''. The source brightness distribution is modelled as a collection of point sources, which are extracted at the location of the peaks in the dirty image through an iterative process until some specified stopping criteria is reached. In observations with limited \uv sampling, such as those obtained with the EHT, it is important to guide the \texttt{CLEAN} deconvolution process through the inclusion of the so-called ``cleaning windows'', restricting the sky areas within which the point components are localized.

Mitigation of the \textit{a priori} calibration uncertainties is commonly carried out through multiple rounds of \texttt{CLEAN} deconvolution followed by self-calibration, which solves for the station gains that maximize consistency between the current model and the measured visibilities \cite[e.g.,][]{Wilkinson1977, Readhead1980, Cornwell1981, Pearson1984}. Amplitude self-calibrations are necessarily limited to intervals of time larger than the expected variability in order to retain information about source variability. The final image is obtained by convolving the model components with a Gaussian CLEAN beam that approximates the central lobe of the point-spread function of the interferometer, with the addition of the last residual image, which  represents some unconvolved additional structure and noise. In this paper we use the \texttt{Difmap} software  \cite[e.g.,][]{shepherd1997a, shepherd_2011} for \texttt{CLEAN} imaging.

Once the imaging procedure converges based on a specified stopping criterion into an average static image, \texttt{CLEAN} dynamic imaging is performed by first dividing the data set into smaller portions with a time duration similar to that of the expected source time scale variability (i.e., ``snapshots''). Under the assumption of small structural changes over time, the model corresponding to the static image is used as an initial model, upon which we look for structural changes by cleaning the residual map corresponding to each data snapshot. To guide the deconvolution with such a limited \uv coverage, we limit the extra cleaning to the imaging regions in which we have emission in the averaged image by placing tight cleaning windows. In addition, further self-calibrations in phase and amplitude are performed to refine antenna gain corrections.
  
The \textrm{CLEAN} algorithms do not enforce similarity between snapshots, other than the use of common initial image priors, which facilitates tracking of rapid source structural changes at arbitrarily separated spatial locations. However, these image changes are restricted to occur within the tight cleaning windows established around the emission found in the averaged static image.

\subsection{Forward-modeling approaches}
\label{sec:rml}

Unlike the inverse-modeling methods, which solve for a sparse image on the image domain from the dirty map transformed from the measurement sets, the forward modeling methods solve for an image by evaluating the data likelihood derived from the consistency between actual measurements and the model dataset forward-transformed from the image.
It offers flexibility to the imaging through robust data products (e.g., closure quantities that are not affected by station-based calibration uncertainties) and incorporates various observing effects into the observational equation used in the forward transform.

Regularized Maximum Likelihood (RML) methods \citep[see][for an overview]{EHT4} optimize a cost function composed of $\chi^2$ terms (proportional to log-likelihood terms) of visibility components and regularization terms that describe the prior assumptions for images. Each regularization term is described by a product of its relative weight (i.e. hyperparameter) and regularization functions. These regularization functions include, e.g., maximum entropy \citep[e.g.][]{Narayan1986, Chael_2016}, total variation and its variants \citep[e.g.][]{Akiyama_2017a, Kuramochi2018}, and sparsity priors \citep[e.g.][]{Honma_2014}. The cost function can be interpreted as a maximum-a-posteriori (MAP) estimation by considering the regularization terms as log prior distribution of the image, although regularization functions do not always have a probabilistic interpretation. The final reconstruction is convolved with the CLEAN beam of the interferometer to remove the effects of methodology-specific super-resolution.

The RML approach can be extended to dynamic reconstruction (henceforth RML dynamic imaging) in a conceptually simple way \citep[][]{Johnson2017}. 
The likelihood term can be formulated by forward-transforming snapshots of a video, instead of a single image, to data.
One can add temporal regularization terms, that penalize temporal variations of the source structure by defining a metric for the ``distance'' between adjacent frames. 
A popular choice is a sum of squared pixel differences between two adjacent snapshots, assuming that snapshot-to-snapshot transition of the source brightness is piecewise smooth (e.g. the $R_{\Delta t}$ regularizer in \citealt{Johnson2017}).
Another widely-used choice is a sum of squared differences between the time-averaged image and each snapshot, based on an assumption conceptually similar to dynamic \texttt{CLEAN} imaging (\autoref{sec:clean}) that the deviations of each snapshot from the mean image is small and sparse  (e.g. $R_{\Delta I}$ regularizer in \citealt{Johnson2017}). The temporal regularization term necessarily suppresses intrinsic source variability if weighted too high; however, what constitutes ``too high'' varies depending on the source structure and variability timescale.
Popular image distance metrics include the Euclidean norm or a relative entropy such as Kullback–Leibler divergence \citep{kullback_1951}.

StarWarps \citep{Bouman2018} is another forward-modeling method for dynamical imaging adopted in this work. 
StarWarps, based on a probabilistic graphical model, solves for snapshots of a video by solving its posterior probability distribution defined as a product of three terms: data likelihood, multi-variate Gaussian distributions for each snapshot, and transitional probabilities between adjacent snapshots effectively working as spatial and temporal regularizations, respectively. 
StarWarps allows for the exact inference of the video by computing a mean and covariance of the image, which provides a complete description under Gaussian approximation. By contrast, the RML dynamic reconstruction derives only a MAP estimation. StarWarps requires an initial static image, which can either be data-driven (e.g., a best-fitting static image of the entire dataset being dynamically reconstructed) or prior-driven (e.g., a synthetic image of a ring).

In this paper, we use the RML dynamic imaging algorithms implemented in \texttt{eht-imaging} (also referred to as \texttt{ehtim}) and \texttt{SMILI} (Sparse Modeling Imaging Library for Interferometry), and the StarWarps algorithm in \texttt{eht-imaging}. See \textbf{EHT Sgr A* Paper III} for more details of regularization functions and other imaging parameters used in the reconstructions.

\section{Limitations of sparse \uv coverage}
\label{sec:demo}

In this section, we explore limitations of imaging with limited and directionally-biased \uv coverage. In particular, we will show that sparse \uv coverage results in predictable limitations (i.e., deviations from the true source morphology), which are determined by the geometry of the \uv coverage. 

\begin{figure}[h]
\centering
\includegraphics[width=\columnwidth]{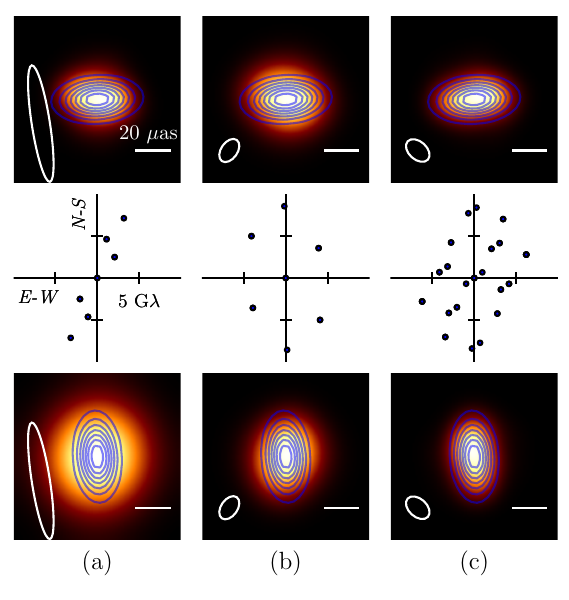}
\caption{Images from a snapshot RML reconstruction of a simulated 2017 EHT observation of a rotating elliptical Gaussian. The reconstructions, shown in orange, are compared to the model images, shown in blue contours corresponding to 5\%, 20\%, 40\%, 60\%, and 80\% of the model maximum brightness. The top row of panels are reconstructions of an approximately horizontally-oriented elliptical Gaussian, while the bottom row of panels are reconstructions of an approximately vertically-oriented Gaussian. The bottom left ellipse in each image subfigure shows the elliptical CLEAN beam. The middle row of panels show the snapshot coverages used to produce the reconstructions shown. The coverage shown in (a) is produced by the simultaneous observing of ALMA, APEX, SPT, and LMT; in (b), ALMA, APEX, SPT, and PV; in (c), all sites except for PV. Snapshots (a) and (b) have an identical number of baselines in their \uv coverage configuration but different angular isotropy. The \uv coverage in (b) is relatively capable of resolving the source along the North-South and East-West baselines, and orients the reconstruction properly. By comparison, the \uv coverage in (a) can only constrain flux in the direction of the collimated coverage. As a result, it incorrectly recovers the orientation of the Gaussian. The \uv coverage in column (c) is a maximally-\uv-filling case for the 2017 array configuration, with baselines relatively evenly distributed both radially and angularly. The result is an accurate recovery of the model behavior.
} 
\label{fig:vuln_pedagogical}
\end{figure}

\begin{figure}[h]
\centering
\includegraphics[width=\columnwidth]{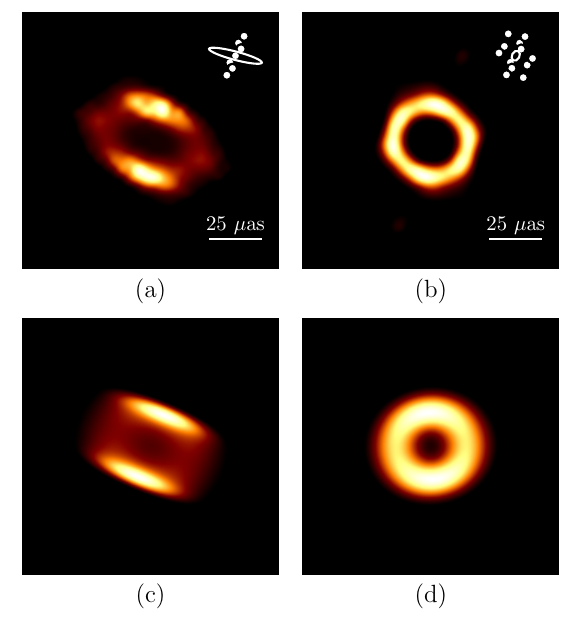}
\caption{Images from a snapshot reconstruction of a simulated 2017 EHT observation of an axisymmetric static ring. Panel (a) is chosen to represent a portion of the observation where the \uv coverage was heavily anisotropic, and panel (b) is chosen to represent a portion of the observation where the \uv coverage was approximately isotropic and radially dense. The reconstruction is demonstrably and predictably affected by the angular homogeneity of the instantaneous \uv coverage, shown as a set of white dots. The white ellipse corresponds to the CLEAN beam and is directly linked to the \uv coverage. The predictability of the image artifacts resulting from \uv coverage is exemplified by the bottom panels, which show the model ring convolved with the CLEAN beams shown in the top row. Simple convolution with the CLEAN beam is enough to reproduce the salient artifacts.
}
\label{fig:asymmetry_ring_reco}
\end{figure}

\autoref{fig:vuln_pedagogical} shows how imaging using directionally-biased \uv coverage (column A) fails to properly recover the orientation of an intrinsically non-circularly-symmetric source when the \uv coverage does not sufficiently sample the source structure in the relevant direction. 
By contrast, the same baselines--oriented in a more isotropic way (column B)--are capable of recovering the source profile in all directions. In addition, reconstructions of circularly symmetric sources by directionally-biased \uv coverage (column A of \autoref{fig:asymmetry_ring_reco}) can introduce a lack of circular symmetry that is not present in the underlying source or in reconstructions performed on non-directionally-biased coverage (column B of \autoref{fig:asymmetry_ring_reco}). Due to the minimization of the artifacts introduced into a reconstruction via incomplete \uv coverage, imaging algorithms work better when applied to isotropic \uv coverages.

\begin{figure}[ht]
\centering
\includegraphics[width=\columnwidth]{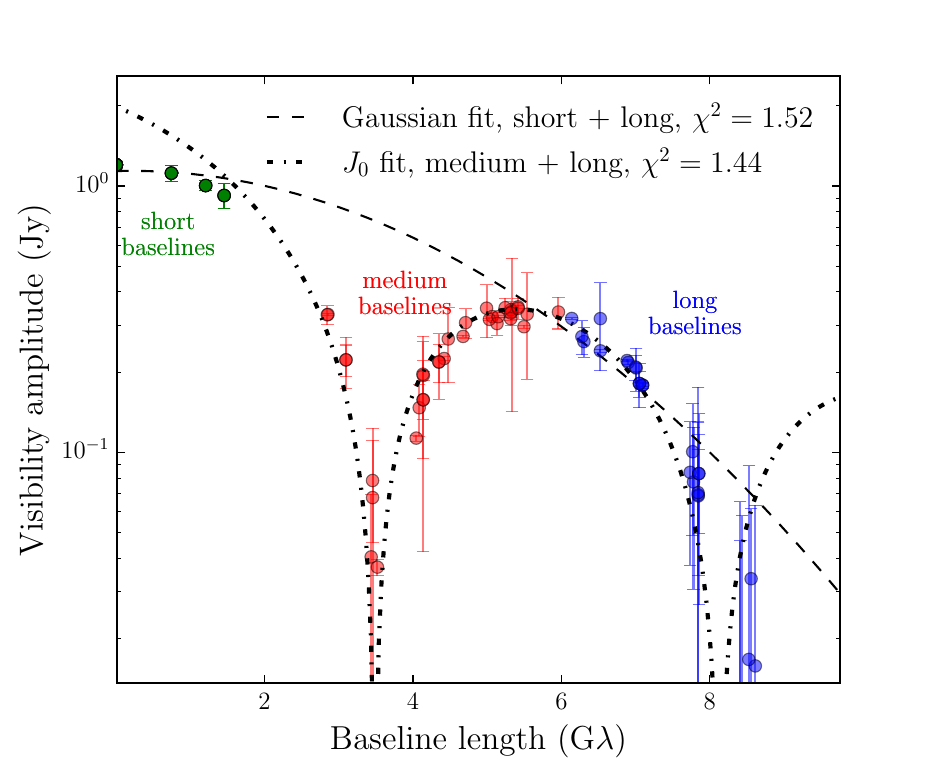}
\caption{Visibility amplitudes of a simulated 2017 EHT observation of a ring (25 $\mu$as radius, 5 $\mu$as width, convolved with 10 $\mu$as circular Gaussian) are shown as a function of radial baseline length $\rho$. Short ($\leq 2$ G$\lambda$), medium ($2$ G$\lambda$ $< \rho < 6$ G$\lambda$), and long ($\geq 6$ G$\lambda$) baselines are displayed in green, red and blue, respectively. Continuous fits of different (but equally well-fitting) models are overlaid. Fitting a ring model with infinitesimal thickness (denoted by $J_0$, representing a zeroth-order Bessel function of the first kind) to only medium and long baselines accurately represents the source shape and size, but poorly constrains the total intensity of 1 Jy. In addition, an equally good fit can be obtained with a simple Gaussian fit to only short and long baselines. Data from all three baseline types are required to correctly constrain key properties of the source. This result highlights how model misspecification can lead to severe systematic errors, especially when working with limited baseline coverage. }
\label{fig:vuln_radial}
\end{figure}

In addition to angular inhomogeneity, radially inhomogeneous coverage also leads to ambiguous image reconstruction. The Fourier transforms of various source types (e.g., rings, crescents, Gaussians, etc.) are approximately degenerate when observed using an interferometer that only marginally resolves the image \citep[e.g.,][]{TMS, Issaoun2019}. 
More complex approximate degeneracies exist for radially inhomogeneous interferometers which only probe short and long but not intermediate baselines \citep[e.g.,][]{Doeleman2008}. For the EHT observing a ${\sim}50\ \mu$as diameter ring, as expected for \sgra, short baselines correspond to those with length $\textless 2$ G$\lambda$ and long baselines correspond to those with length $\textgreater 6$ G$\lambda$. \autoref{fig:vuln_radial} demonstrates that a Gaussian model describes the simulated EHT observation of a 50 $\mu$as static ring nearly as well as the static ring model itself if only particular subsets of the data are fit. Even an infinitesimally thin ring model, when only fit to medium and long baselines, can provide a high-quality fit while misrepresenting the total intensity of the source.
Without sufficient radial homogeneity (i.e., coverage of short, medium, and long baselines, as seen in column C of
\autoref{fig:vuln_pedagogical}), fitting and interpreting a model confidently can be difficult. 

Periods of \uv coverage where these limitations are more likely to occur can be identified by constructing a metric that scores directional bias and radial homogeneity (i.e., coverage of short, medium, and long baselines). The prevalence and severity of reconstruction artifacts that result from coverage limitations form a continuum that can be used to rank different \uv configurations. A metric based on these limitations could be applied to a full observation to distinguish different observing periods (composed of many evolving \uv configurations) by their ability to produce high-quality reconstructions.

\section{Coverage metrics}
\label{sec:alternatives}

Multiple \uv coverage metrics with different underlying considerations exist in the literature. Here, we summarize several metrics and compare the way they score a given observation. In addition, we develop a novel ``isotropy metric'' that has been tailored to the specific vulnerabilities detailed in \autoref{sec:demo}.

Our ``selective dynamical imaging'' approach uses such a metric to identify intervals during an observation where the coverage is optimally configured for imaging. Importantly, these intervals are chosen before image reconstruction is attempted. In contrast, methods such as “lucky imaging” (e.g., \citealt{Fried1978}) identify particularly useful images (e.g., with minimal distortion) after and on the basis of the reconstruction. A single score computed directly from \uv coverage known \textit{a priori} is preferable to an empirical approach (e.g., performing a simulated observation of a synthetic model and comparing reconstructions with the model) as it provides source-structure-agnostic assessments with a substantially lower performance cost.

Any metric capable of scoring different periods of \uv
coverage would have demonstrable limitations. Within a single observation, a comparison of \uv coverage at two different points in time is a reliable way of determining which time region of the observation will produce superior image reconstructions. However, certain reconstruction-impacting data quantities can vary independently of the \uv coverage. Sensitivity, calibration, and systematic uncertainty can also be important factors, but are not probed by coverage metrics. 

\subsection{Normalized cross-correlation}
\label{sec:nxcorr}

The normalized cross-correlation between two images is a measure of their similarity. By performing a dynamical reconstruction on a simulated observation and comparing each image of the dynamical reconstruction to the model, we can heuristically identify which portions of the observation produced the best reconstruction (i.e., the portions of the reconstruction with the greatest similarity to the model). We define the normalized cross-correlation $\rho_{\textrm{NX}}(X,Y)$ of two images $X$ and $Y$ in an identical fashion to \cite{EHT4},

\begin{equation}
    \rho_{\textrm{NX}}(X,Y) = \frac{1}{N}\sum_i\frac{(X_i-\langle X \rangle)(Y_i-\langle Y \rangle)}{\sigma_X \sigma_Y}.
    \label{eq:nxcorr_definition}
\end{equation}

\noindent To compute the normalized cross-correlations used in \autoref{fig:all_metrics}, we reconstruct a static ring (one image per snapshot of the observation) and compute the metric value between the model image and the reconstruction. 

A normalized cross-correlation between a model and the associated reconstruction is the most straightforward way to identify trustworthy periods of an observation, assuming that the reconstruction on synthetic data will behave in a similar fashion to a reconstruction on real data. This assumption is only upheld if care is taken to ensure that the qualities of the synthetic data match those of the real data.
In addition, heuristic tests such as the normalized cross-correlation can be biased depending on the structure and inherent variability of the model chosen. If the source randomly aligns with collimated \uv coverage at some point in the observation, it can result in a misleadingly high normalized cross-correlation that cannot be replicated for a different source model.

\subsection{\uv filling fraction}

\citet{Palumbo_2019} proposes a geometric scoring procedure for the \uv coverage based on the specification of a desired array resolution $\theta_{\rm res}$ and imaging field of view $\theta_{\rm FOV}$. $\theta_{\rm res}$ sets an outer boundary with radius $1/\theta_{\rm res}$ in the \uv plane within which a ``filling fraction'' is computed, and is typically taken to be the nominal array resolution set by the longest baseline in an observation. $\theta_{\rm FOV}$ determines a convolution radius of $0.71/\theta_{\rm FOV}$ corresponding to the scale in the \uv plane over which the Fourier response to a filled disc on the sky of diameter $\theta_{\rm FOV}$ would decay to half of its maximum amplitude; we use $\theta_{\rm FOV}=100{\rm \mu as}$ for the filling fraction computation in  \autoref{fig:all_metrics}. Intuitively, the largest image feature considered in the optimization of coverage sets the smallest scale of interest in the \uv plane; thus, convolving a proposed set of \uv points by $0.71/\theta_{\rm FOV}$ yields a measure of what region in the \uv plane is sampled by measured visibilities. The fraction of the bounding circle sampled by the convolved coverage is the filling fraction.

Increasing the specified resolution (perhaps by increasing observation frequency) extends the bounding circle, decreasing the filling fraction unless $\theta_{\rm FOV}$ is correspondingly decreased. 
In this way, the filling fraction captures some features of the ``spatial dynamic range'' discussed in \citet{Lal_2010}. As shown in figure 7 of \citet{Palumbo_2019}, the filling fraction metric is a successful and nearly linear predictor of image fidelity until the filling fraction reaches values near 0.9, at which point imaging techniques are limited by methodology-specific super-resolving scales, which for many imaging algorithms is at approximately half of the diffraction-limited CLEAN beam width in the case of the EHT \citep{EHT4}.

\subsection{Largest coverage gap}
\label{sec:lgc}
An alternative metric probing the coverage isotropy is based on identifying the largest gap in the \uv coverage, hence we refer to it as the largest coverage gap (LCG) metric \citep{github_LCG}.
In this approach we consider the coverage as a set of sampled \uv plane locations, to find the largest circle that can be drawn within the limits of the coverage that does not contain a coverage point in itself. Such a largest circular gap can be efficiently calculated with Delaunay triangulation of the coverage set \citep{Barber1996}. Then the diameter of the gap $d_{\rm max}$ can be turned into a metric coefficient with
\begin{equation}
    m_{\rm LCG} = 1 - d_{\rm max}/\rho_{\rm max} \ ,
    \label{eq:lcg}
\end{equation}
where $\rho_{\rm max}$ is the longest projected baseline length. If we demand that the $(u,v)$ distance corresponding to the center of the circle is less than $\rho_{\rm max}$, then we have $0\le  m_{\rm LCG} \le 1 $, with  $m_{\rm LCG}=1$ corresponding to the limit of a complete continuous coverage. A coverage consisting of a single detection would correspond to  $m_{\rm LCG} = 0$. Unlike the filling fraction metric, the LCG metric is independent of the assumed field of view.

\subsection{Isotropy and radial homogeneity}
\label{sec:isotropy}
We propose a novel metric of \uv coverage isotropy and radial homogeneity (hereafter referred to as the ``isotropy metric'') based on the limitations described in \autoref{sec:demo}. Similarly to the LCG metric, the isotropy metric penalizes anisotropy of the coverage, although the two approaches differ appreciably. In this approach, we treat the distribution of baselines in the Fourier plane as a mass distribution and quantify the radial and angular homogeneity using the second moments of inertia. We define the isotropy metric coverage parameter $\mathcal{C}$ for a given snapshot as
\begin{align}
    &\mathcal{C} = I\left(1-\frac{\mathcal{K}}{\mathcal{K}_{\textrm{max}}}\right),
\end{align}
with 
\begin{align}
    I = 1-\frac{\sqrt{(\langle u^2 \rangle-\langle v^2 \rangle)^2+4\langle uv \rangle^2}}{\langle u^2 \rangle+\langle v^2 \rangle},
\end{align}
where $\langle u^2 \rangle$, $\langle v^2 \rangle$ and $\langle uv \rangle^2$ are the second moments of the baseline distribution, $\mathcal{K}$ is the Kolmogorov-Smirnov (KS) distance of the radial distribution of baseline lengths from uniform, and $\mathcal{K}_{\textrm{max}}$ is the maximum value of $\mathcal{K}$ at any point in time during the observation (or an arbitrary value for the purpose of cross-observation comparisons). The isotropy metric has the benefit of being fully analytic and automatically normalized between 0 and 1. A full derivation of the isotropy metric is presented in \autoref{sec:derivation}.


\subsection{Discussion of metrics}

Despite differences in methodology and implementation, the metrics we examined found similar fluctuations in \uv coverage quality and identify similar candidate time regions for high-quality imaging. A comparison of the metrics detailed in Subsections \ref{sec:nxcorr}-\ref{sec:isotropy} applied to the 2017 EHT \uv coverage of \sgra is shown in \autoref{fig:all_metrics}. In general, the second half of the observation has superior \uv coverage as indicated by the metrics, and the period from ${\sim}$01:00 GMST to ${\sim}$03:30 GMST maximizes the various metrics.

The \uv filling fraction and LCG metric generally produce results similar to the isotropy metric. Disagreements between the metrics can be seen especially at the beginning of the observation (e.g., 17--19 GMST) and in the middle (e.g., 21--23 GMST). These are periods where the \uv coverage is extremely sparse and not suitable for imaging, though the degree to which these periods are determined to be unsuitable varies depending on the specific considerations of the individual metrics. 

The normalized cross-correlation metric, while the most direct measurement of time-varying reconstruction quality, lacks source-structure agnosticism. If the source structure is known, the normalized cross-correlation metric can be a useful method of determining what periods of an observation are most advantageous for imaging that particular source structure. However, if the source structure is unknown, then a wide suite of representative source models must be tested to mitigate possible biases. Additionally, the normalized cross-correlation method will be tied to the particular imaging algorithm and hyperparameters used, making this metric less robust than the others considered.

The particular constructions of the metrics can lead directly to unintuitive or undesirable behavior. One example of undesirable behavior is a metric punishing a coverage for adding data points. Intuitively, more baselines lead to better coverage of the Fourier plane and therefore more information about the source. However, if these additional baselines are placed strategically, they can result in an unintuitive score assignment. A trivial example of this can be generated for the LCG metric. Consider a coverage with maximum baseline length less than $\rho$ that achieves  $m_{\textrm{LCG}}{\approx}1$. By placing a single baseline of length $L$ far outside the initial coverage (i.e., $L\gg \rho$), $\rho_{\textrm{max}}=L$ and $d_{\textrm{max}}$ goes as ${\sim}L-\rho$. This drives $m_{\textrm{LCG}}$ to zero and seems to indicate the coverage has become demonstrably worse, when in reality the coverage quality has largely stayed the same, with the improvement of a single ultra-long-baseline. This type of array pathology does not occur in the 2017 EHT coverage but may present an issue if the EHT goes to ultra-long space baselines.

\begin{figure}[h]
\centering
\includegraphics[width=\columnwidth]{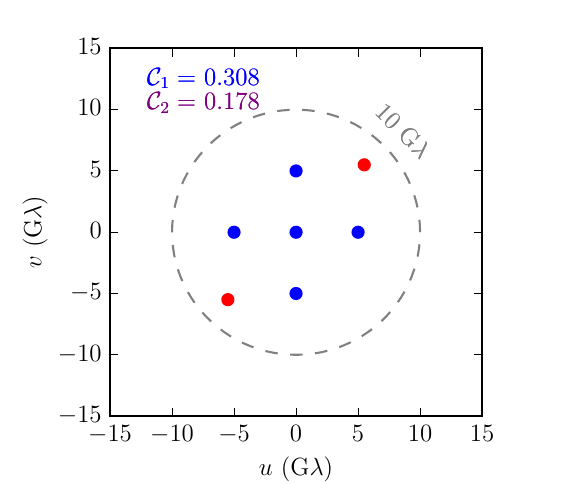}
\caption{A pathological case for the isotropy metric which demonstrates an undesirable behavior. The isotropic coverage shown (blue) results in an isotropy metric value of ${\approx}$0.3 ($\mathcal{C}_1$, shown in blue). Adding two data points (red) strategically (i.e., in an anisotropic configuration) decreases the overall isotropy of the array and lowers the metric score by a factor of ${\approx}1/2$ ($\mathcal{C}_2$, shown in red+blue=purple). This change makes sense given the considerations of the metric--the new array is more anisotropic and therefore has a lower score. However, this behavior is undesirable since, intuitively, we expect that an array with more baselines will perform better than an array with fewer baselines. Note: in order to compute the isotropy metric as defined in \autoref{sec:isotropy}, the example coverages shown above are assumed to be part of the Apr 7, 2017 EHT coverage of \sgra and the corresponding value of $\mathcal{K}_{\textrm{max}}$ is adopted (see \autoref{sec:derivation}).}
\label{fig:sdi_isotropy_limitation}
\end{figure}

The isotropy metric exhibits similar misbehavior, as demonstrated in \autoref{fig:sdi_isotropy_limitation}. Given an isotropic coverage with a low number of baselines, adding just two baselines strategically can decrease the isotropy metric value by a substantial amount. With so few baselines, the addition of new baselines would intuitively be considered an improvement. However, the metric detects a decrease in isotropy and reports accordingly. This problem is only present for arrays with small numbers of baselines--it is difficult or impossible to significantly alter the isotropy of an array configuration for larger arrays using only a few baselines without resorting to ultra-long-baseline placement as in the LCG example.

An additional limitation that any metric based purely on coverage possesses results from unusual source structure. The metrics described above attempt to predict reconstruction quality by analyzing the coverage available, but this prediction is performed under the assumption of ``reasonable'' source structure (i.e., source structure with smooth, continuous Fourier representation). However, we can construct simple examples that would render these metrics unhelpful by violating the assumption of reasonable source structure. Consider a source whose Fourier transform has zero flux density everywhere an array has \uv coverage, and non-zero flux density everywhere else. Regardless of how good the coverage itself is (and therefore how well a given metric may score the coverage), there is no way to produce an accurate reconstruction of the source. This limitation is not likely to be an issue, as the restriction that source emission be non-negative induces sufficient correlation in the Fourier domain that the possibility for arbitrary pathologies to ``hide'' in coverage-deficient swathes of the plane is severely limited.

The isotropy metric offers substantial performance (i.e., time and overhead cost) benefits over the other metrics considered, while reporting similar results. We test the performance of each metric by computing the per-snapshot score for a full 12 hour observation five times, resulting in $\sim$10k snapshots available for scoring. We run the performance assessments on a \texttt{i5-1038NG7} 10th-generation Intel \texttt{x86\_64} processor with 16 GB of RAM. The normalized cross-correlation, which must reconstruct hundreds of images and compare them to model images, takes several hours to complete. The isotropy metric performs ${\sim}10^5$ times faster than the full normalized cross-correlation, ${\sim}20$ times faster than the \uv filling fraction and ${\sim}10$ times faster than the LCG metric, while producing similar assessments of coverage. The substantial performance differences between the metrics will become more pronounced with larger arrays, such as the next-generation EHT (ngEHT) coverage. The low overhead generated by the isotropy metric in comparison to the other metrics examined makes the isotropy metric a more optimized method of scoring \uv coverages in important contexts, such as real-time track selection and long-term ngEHT site placement \citep{Raymond_2021}, which is expanded upon in \autoref{sec:inter-day}.

\section{Application of metric to 2017 EHT array}
\label{sec:application_2017}

\begin{figure*}[t]
\centering
\includegraphics[width=\textwidth]{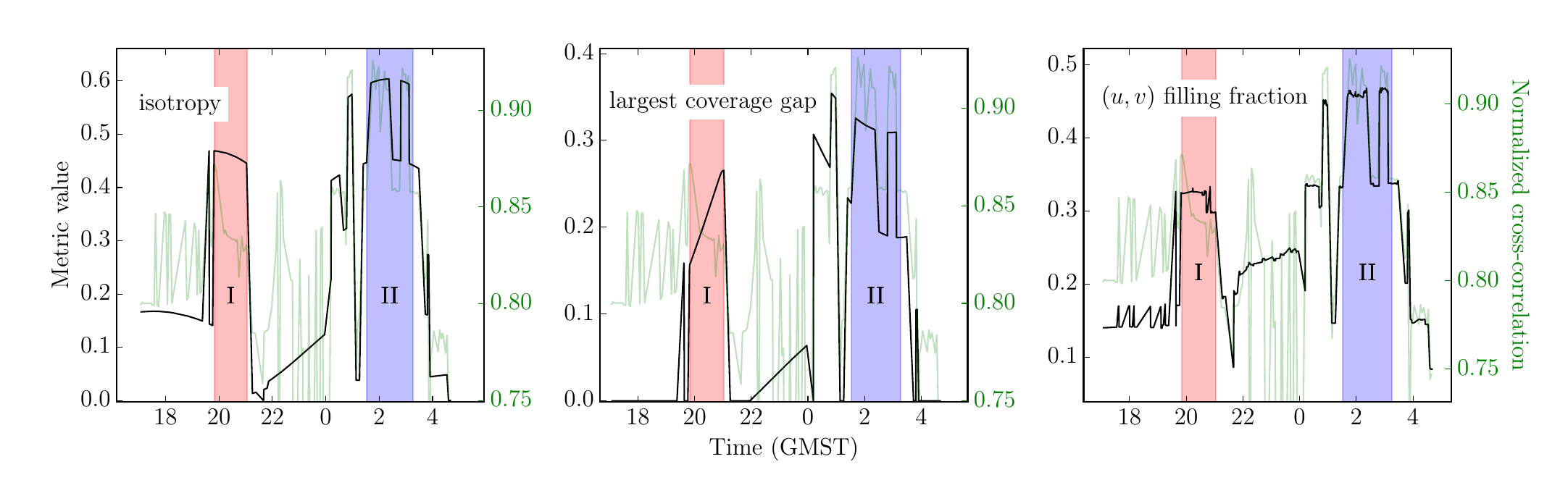}
\includegraphics[width=0.65\textwidth]{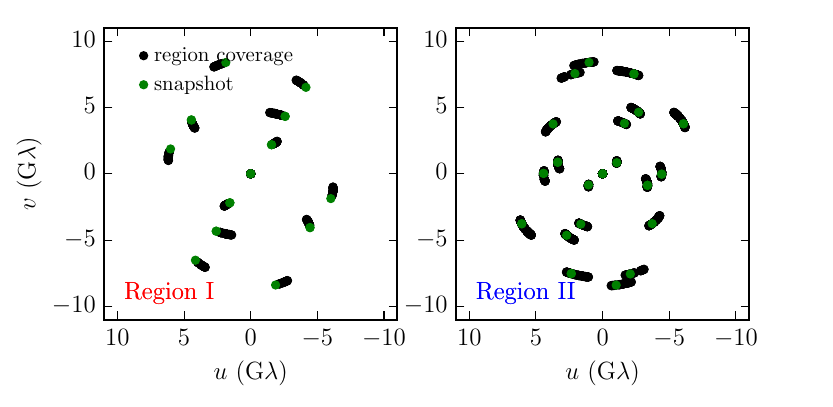}
\caption{The application of all metrics (top left: isotropy; top middle: largest coverage gap; top right: \uv filling fraction) described in \autoref{sec:alternatives} to the 2017 EHT coverage of \sgra. The observation begins on April 7, and the 00:00 GMST indicates the day change to April 8. The red time region corresponds to Region I, and the blue time region to Region II, as described in \autoref{sec:application_2017}. The normalized cross-correlation (green, representing the comparison between a snapshot reconstruction of a static ring and the model) is shown as a reference for all metrics as it is the only direct measure of instantaneous reconstruction quality. All three \uv-coverage-based metrics show high-quality coverage in the time region from ${\sim}$01:30 GMST to ${\sim}$03:10 GMST (Region II). The bottom row of the figure shows the coverage of Region I (bottom right) and Region II (bottom left) along with a representative snapshot.}
\label{fig:all_metrics}
\end{figure*}

We apply the coverage metrics discussed in \autoref{sec:alternatives} to the EHT \uv coverage of \sgra corresponding to April 7, 2017. The instantaneous metric values for each snapshot of the observation are shown in \autoref{fig:all_metrics}. This method of scoring the observation clearly identifies distinct periods of varying coverage quality. The time region from ${\sim}$01:30 Greenwich Mean Standard Time (GMST) to ${\sim}$03:10 GMST (denoted as ``Region II'' in \autoref{fig:all_metrics}) has the highest overall isotropy and baseline density of the observation. We select this period as a candidate time region for high-quality imaging (a ``good'' time region, i.e., one where the typical metric score is high). In contrast, the time region from ${\sim}$19:45 GMST to ${\sim}$21:00 GMST (denoted as ``Region I'' in \autoref{fig:all_metrics}), while relatively stable, displays substantially lower coverage quality. We select this time region to examine the behavior of reconstructions in periods of ambiguous coverage quality. Exact timestamps for these time regions are given in \autoref{tab:timestamps}.

\begin{table}[]
               \aboverulesep=0ex 
               \belowrulesep=0ex 
\centering
\begin{tabular}{@{}lll|ll@{}}
            & \multicolumn{2}{c}{\textbf{Start time}}              & \multicolumn{2}{c}{\textbf{End time}}  \\
            \cmidrule(lr){2-3}
            \cmidrule(lr){4-5}
            & \multicolumn{1}{c}{UT}          & \multicolumn{1}{c}{GMST}        & \multicolumn{1}{c}{UT}  & \multicolumn{1}{c}{GMST}        \\ \toprule

Observation   & 4.0458     & 17.087   & 15.598    & 4.6721 Apr 8  \\ \greyrule
{\color{red}Region I}    & 6.7766     & 19.825  & 7.9763    & 21.028  \\ \greyrule
{\color{blue}Region II}   & 12.618 & 1.6834 Apr 8  & 14.046   & 3.1155 Apr 8  \\ 
\textit{LMT drop} & 13.435  & 2.5045 Apr 8  & 13.737  & 2.8058 Apr 8  \\ \bottomrule
\end{tabular}
\caption{Timestamps for the observation and beginning and end of the time regions of interest in the Apr 7, 2017 EHT coverage of \sgra. Region I and Region II correspond to the time regions identified in red and blue (respectively) in \autoref{fig:all_metrics}. The LMT dropout corresponds to the sudden loss of coverage that occurs partway through Region II. All timestamps correspond to Apr 7, 2017 unless otherwise noted. In UT, the observation begins and ends on Apr 7; however, when converted to GMST, Region I lies on Apr 7, 2017 while Region II lies on Apr 8, 2017.}
\label{tab:timestamps}
\end{table}

\begin{figure*}[h]
\centering
\includegraphics[width=0.825\textwidth]{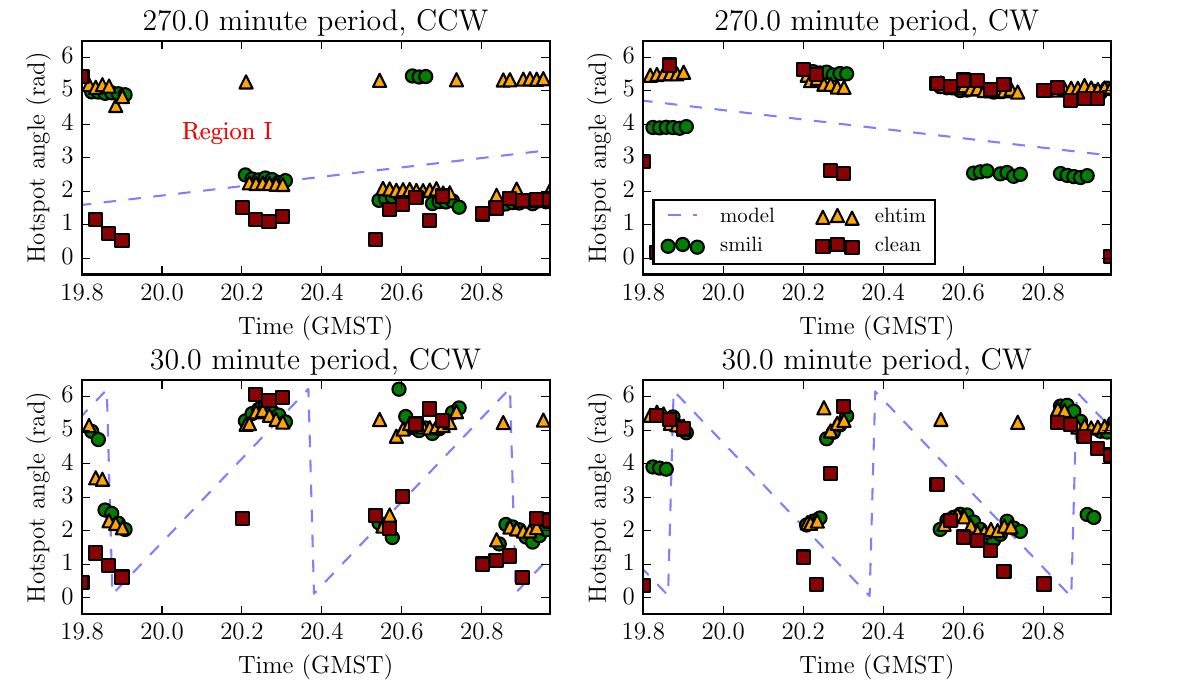}
\caption{\texttt{SMILI} (green circles), \texttt{ehtim} (yellow triangles) and \texttt{CLEAN} (red squares) position-angle reconstructions in Region I of a \texttt{ring+hs} model for orbits of various periodicities and directions.  Comparing the resulting hotspot position angle to the model angle (blue dashed line) clearly demonstrates that Region I provides a poor reconstruction of time variability.} 
\label{fig:ehtim_reco_hotspot_orientation_regionI}
\end{figure*}

\begin{figure*}[h]
\centering
\includegraphics[width=0.825\textwidth]{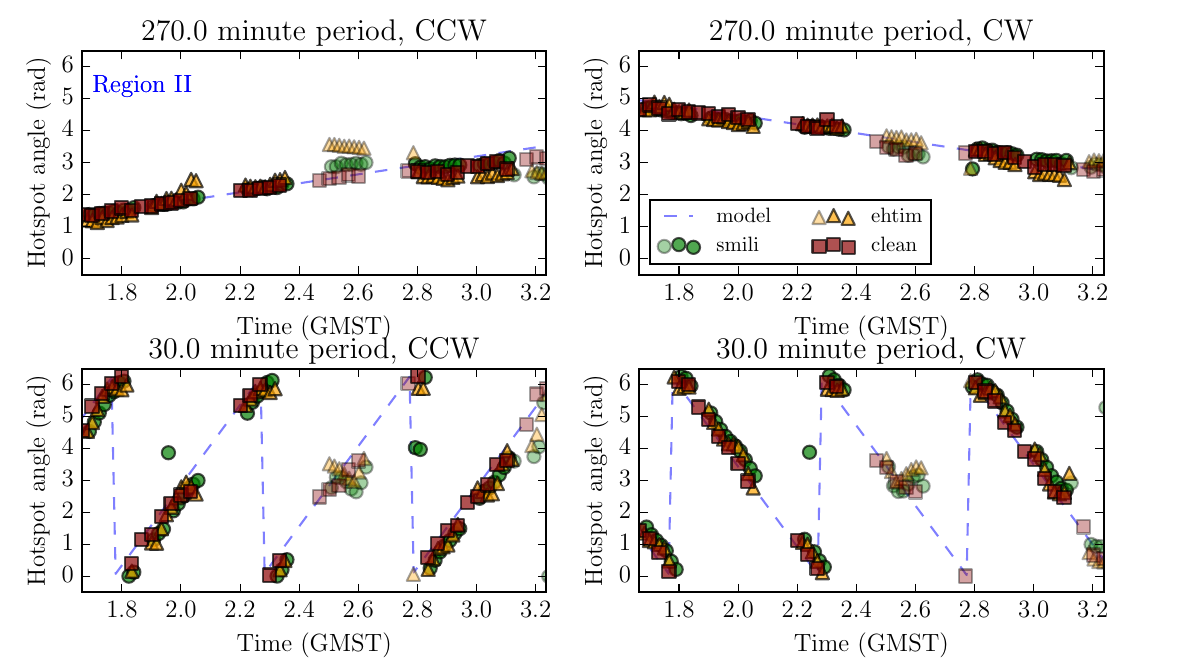}
\caption{\texttt{SMILI} (green circles), \texttt{ehtim} (yellow triangles) and \texttt{CLEAN} (red squares) position-angle reconstructions in Region II of a \texttt{ring+hs} model for orbits of various periodicities and directions. The transparency of each data point corresponds to the isotropy of the snapshot observation used to reconstruct it. Comparing the resulting hotspot position angle to the model angle (blue dashed line) clearly demonstrates that Region II provides excellent conditions and coverage for meaningful recovery of periodicity given reasonable imaging assumptions.} 
\label{fig:ehtim_reco_hotspot_orientation_regionII}
\end{figure*}

By performing reconstructions in these time periods, we can validate the capability of the \uv coverage metrics to predict reconstruction quality based on coverage alone. We reconstruct
four configurations of the \texttt{ring+hs} toy model detailed in \autoref{sec:model_data}: a 270 minute orbital period clockwise (CW) and counterclockwise (CCW), and a 30 minute orbital period clockwise and counterclockwise. The reconstructions in each time region are produced according to the RML and \texttt{CLEAN} imaging methods in \autoref{sec:imaging}, and we perform feature extraction on each image using the \texttt{REX} module of \texttt{eht-imaging} (\citealt{EHT4}) to recover the position angle of the hotspot at each moment in time. The extracted hotspot angles and the model orbits for Regions I and II are shown in Figures \ref{fig:ehtim_reco_hotspot_orientation_regionI} and \ref{fig:ehtim_reco_hotspot_orientation_regionII}, respectively. Images sampled from dynamical reconstructions generated by each method are displayed in \autoref{sec:appendix_sampled_results}. The ``success'' of a reconstruction is determined by the successful extraction of the hotspot position angle.

We find that Region II produces reconstructions that facilitate accurate recovery of dynamical variability. The reconstructions show a ring of approximately $50 \ \mu$as in diameter with a distinct hotspot. The recovered hotspot orientations are shown in \autoref{fig:ehtim_reco_hotspot_orientation_regionII}, along with a comparison to the model values. To compare the $N$ recovered position angles $\varphi^r_\gamma$ with the model angle $\varphi^m_\gamma$, we use a phase-adjusted rms $\mathcal{R}$, defined as
\[ \mathcal{R} = \left[\frac{1}{N}\sum_\gamma \begin{cases}
    (\varphi^r_\gamma-\varphi^m_\gamma)^2,& \text{if } |\varphi^r_\gamma-\varphi^m_\gamma|\leq \pi\\
    (2\pi - \varphi^r_\gamma+\varphi^m_\gamma)^2,              & \text{if } |\varphi^r_\gamma-\varphi^m_\gamma|> \pi
\end{cases}\right]^{1/2} \]
Overall, the Region II reconstructions successfully recover the dynamical variability in the model, with rms of the recovered angles and model varying between 0.16 and 0.20 radians. Decreases in coverage quality as measured by the metrics (indicated point-wise in \autoref{fig:ehtim_reco_hotspot_orientation_regionII} by increases in transparency, using the isotropy metric as an example) correlate with lower-quality recovery of the hotspot position angle. These low-quality angle recoveries are most obvious in the \texttt{RML} reconstructions of the counterclockwise $T=30$ minute case. The \texttt{CLEAN} algorithm appears to be more resistant to the sudden loss of coverage, and maintains reconstruction quality even through drops in metric score. Excluding these lapses in coverage quality, Region II clearly facilitates the recovery of source structure and at least one kind of dynamical variability, covering a wide range of periods and directions.

By contrast, a comparatively low-scoring time region (Region I) does not produce reconstructions capable of accurately recovering dynamical variability. Dynamically reconstructed images show a ring-like feature, but the time variation of the brightness asymmetry does not match the model. Overall, the Region I reconstructions fail to recover the dynamical variability in the model, with rms of the recovered angles and model varying between 1.21 and 1.73 radians. The rms on the recovery in Region I is between 7 and 10 times higher than in Region II. For all tests, the scatter around the model is substantially larger than the scatter in Region II, rendering the accurate extraction of a period difficult or impossible. Reconstructions in periods outside of Region I and Region II cannot recover even basic source structure without significant \textit{a priori} information. 

Based on these results, we expect that reconstructions on real data will produce the most accurate and robust recoveries of orbital dynamical variability in Region II of April 7 of the 2017 EHT observations of \sgra. Reasonable imaging and feature extraction procedures failed to produce meaningful results in Region I. This ranking is consistent with the predictions of the coverage metrics described in \autoref{sec:alternatives} and \autoref{sec:derivation}. A wide array of factors impact reconstructions, and the tests presented here do not comprise a realistic synthetic data suite for assessing whether or not Region II can accurately recover dynamical variability for \sgra. The tests provided here solely demonstrate that Region II is the best time region for performing dynamical reconstructions based on coverage considerations. Additional testing on more complex source types with realistic data corruptions can be found in \cite{PaperIII}. 


\section{Metric-based array comparisons}
\label{sec:inter-day}

\begin{figure}[h]
\centering
\includegraphics[width=\columnwidth]{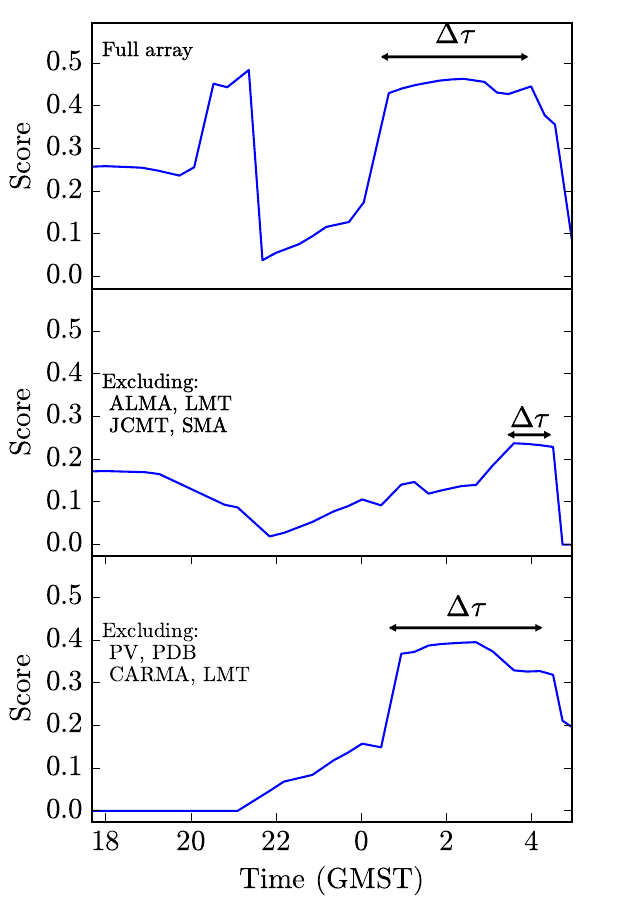}
\caption{Isotropy metric scores for a hypothetical ngEHT array with different dropout scenarios. The full array has all of the sites in the 2017 EHT array (see \autoref{sec:model_data}) with the addition of the Northern Extended Millimeter Array (PDB), the Combined Array for Research in Millimeter-wave Astronomy (CARMA), and the Kitt Peak National Observatory (KP). A universal value of $\mathcal{K}_{\textrm{max}}\approx0.513$ was used to perform the computation (see \autoref{sec:derivation}). Generally, dropouts will impact the maximum coverage score achieved throughout the observation and the duration (shown as $\Delta \tau$) of the most optimal time region. Characterizing a site's importance to the observation is a useful way of informing a go/no-go decision, which takes into account station readiness and probability of dropout. For the above observational scenario, the loss of e.g., ALMA, JCMT and SMA would likely motivate a ``no-go'' decision; whereas the loss of e.g., CARMA and PV would not motivate cancelling the night of observation.}
\label{fig:comparing_tracks}
\end{figure}

Go/no-go decisions about whether to proceed with an observing run are often made with limited information about the readiness or weather conditions at particular sites. Simulating (and scoring) array configurations with different dropouts-- and characterizing changes in the size and quality of the identified candidate temporal regions—can facilitate those go/no-go choices while incorporating uncertainties about station status. One observation configuration can be considered ``better'' than another if it provides a candidate time region with a higher metric score, or a similar metric score for a longer duration.  \autoref{fig:comparing_tracks} shows an example of this kind of analysis performed with a hypothetical ngEHT array. The top panel shows the isotropy metric (see \autoref{sec:isotropy}) score per snapshot for the full array during a night of observation in which every station is observing, which represents the ideal scenario. The middle and bottom panels reproduce the metric score per snapshot assuming 4 sites are unable to observe. By identifying a candidate time region and characterizing its quality (based on, e.g., average metric score in the time region) and duration (denoted as $\Delta \tau$ in \autoref{fig:comparing_tracks}), we can track these characterizations through different combinations of site dropouts and estimate how critical the sites are to the array. Based on the computations in \autoref{fig:comparing_tracks}, ALMA, JCMT, LMT, and SMA dropping out would be catastrophic to the array performance, as the optimal time region for dynamical imaging reduces in duration by $\approx$80\% and reduces in quality by $\approx$50\%. By comparison, the combined dropout of PV, PDB, CARMA, and LMT does not substantially change the duration or quality of the identified candidate time region. 


\begin{figure}[h]
\centering
\includegraphics[width=\columnwidth]{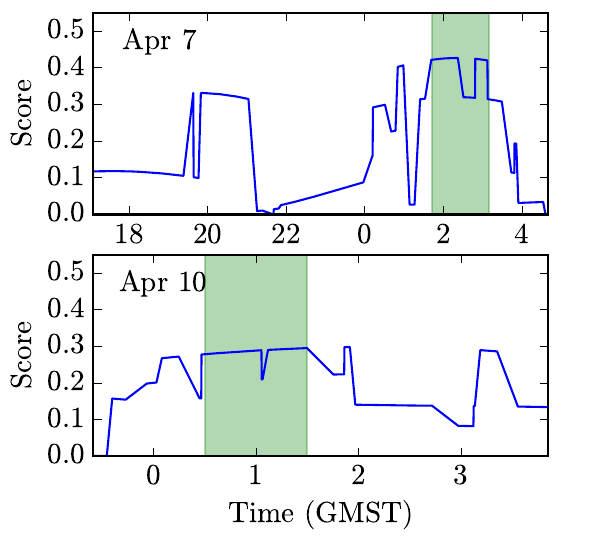}
\caption{Isotropy metric scores for two days of the 2017 EHT observation of \sgra. Candidate time regions for dynamical imaging are highlighted in green. A universal value of $\mathcal{K}_{\textrm{max}}\approx0.513$ was used to perform the computation (see \autoref{sec:derivation}). Dropouts can significantly change the available coverage, resulting in candidate time regions of different duration and quality. By scoring the coverage for the entire observation and directly comparing the predicted quality of candidate time regions, we can identify the best day to perform dynamical imaging. Although the candidate time regions identified on April 7 and April 10 are similar in duration, the candidate time region on April 7 presents $\approx$40\% better coverage.}
\label{fig:comparing_days}
\end{figure}

A metric-based comparison additionally provides a natural and quantitative ranking for identifying which day of an observation campaign produced the most optimal coverage for dynamical imaging. We can rank observations on separate days in a similar fashion to the dropout scenarios visualized in \autoref{fig:comparing_tracks} by comparing the duration and quality of the identified candidate time regions for each day. An example of such a comparison between the Apr 7 and Apr 10 runs of the 2017 EHT observation campaign is shown in \autoref{fig:comparing_days}. Instantaneous metric scores are computed for each scan of both days and candidate time regions for dynamical imaging (green) are identified. While both candidate time regions are of approximately the same duration, the candidate time region associated with Apr 7 displays substantially better coverage, making Apr 7 a better choice for dynamical imaging than Apr 10. 

\section{Conclusions}
\label{sec:conclusions}

We have demonstrated that limitations of imaging associated with sparse baseline distribution can be inferred from the specific geometric properties of the \uv coverage. Highly anisotropic coverage produces artifacts in reconstructions that distort the image in a direction consistent with stripes in the dirty beam. Additionally, an uneven radial distribution of the Fourier
plane (u,v) coverage tends to result in ambiguous image
reconstruction. These limitations can be partially avoided by imaging in radially and angularly homogeneous coverage. The reconstruction issues associated with sparse coverage are exacerbated by rapid short-timescale variability, as seen in a wide array of astrophysical sources, including \sgra and the precessing jets of X-ray binaries. 

Next, we surveyed and compared existing geometric measures of \uv coverage, in addition to deriving a novel isotropy-based metric that addressed the specific limitations demonstrated in \autoref{sec:demo}. The examined metrics included the normalized cross-correlation (\autoref{eq:nxcorr_definition}), the \uv filling fraction (see \citealt{Palumbo_2019}) and the largest coverage gap metric (\autoref{eq:lcg}, see \citealt{github_LCG}). The isotropy metric treats the distribution of baselines as a mass distribution and examines the second moment to rank coverages by homogeneity in the Fourier plane. The isotropy metric gives similar results to other \uv-coverage-based metrics while being more computationally efficient.

These metrics were applied to the April 7 data of the 2017 EHT coverage of Sgr A* (\citealt{PaperI, PaperII, PaperIII, PaperIV, PaperV, PaperVI}) and used to select candidate time regions for high-quality 
dynamical imaging. All metrics identify a period from ${\sim}$01:30 GMST to ${\sim}$03:10 GMST (``Region II'') that minimizes the coverage limitations each metric addresses. We also select a period from ${\sim}$19:45 GMST to ${\sim}$21:00 GMST (``Region I'') with reconstruction capability. Reconstructions of time-variable sources allowed successful recovery of the characteristic source variability in Region II. In contrast, reconstructions in Region I were unable to recover the characteristic motion. The ranking determined by the suite of reconstructions performed on synthetic data verify the predictions made by the examined coverage metrics. We expect that attempts to recover variability in real EHT observations of the Galactic Center will produce the most robust and accurate recoveries in Region II of the April 7, 2017 dataset, and therefore recommend performing dynamical imaging procedures in that time region. 

Coverage metrics have additional utility for ranking inter-observation comparisons based on their ability to recover dynamical variability, which has a variety of applications to the broader field of interferometery. These metrics provide the ability to make select observation time-slots based on the capability of available antennas to recover particular dynamical evolution in the target. Such a scored assessment of coverage could well prove of use to other VLBI arrays both as they make go/no-go decisions about whether to observe on a particular night, and then when identifying the periods of best coverage to perform static and dynamic imaging.

\acknowledgments{We thank the National Science Foundation (awards OISE-1743747, AST-1816420, AST-1716536, AST-1440254, AST-1935980) and the Gordon and Betty Moore Foundation (GBMF-5278) for financial support of this work. This work was supported in part by the Black Hole Initiative, which is funded by grants from the John Templeton Foundation and the Gordon and Betty Moore Foundation to Harvard University. Support for this work was also provided by the NASA Hubble Fellowship grant
HST-HF2-51431.001-A awarded by the Space Telescope
Science Institute, which is operated by the Association
of Universities for Research in Astronomy, Inc.,
for NASA, under contract NAS5-26555.

The Event Horizon Telescope Collaboration thanks the following
organizations and programs: the Academy
of Finland (projects 274477, 284495, 312496, 315721); the Agencia Nacional de Investigación y Desarrollo (ANID), Chile via NCN$19\_058$ (TITANs) and Fondecyt 3190878, the Alexander
von Humboldt Stiftung; an Alfred P. Sloan Research Fellowship;
Allegro, the European ALMA Regional Centre node in the Netherlands, the NL astronomy
research network NOVA and the astronomy institutes of the University of Amsterdam, Leiden University and Radboud University;
the Black Hole Initiative at
Harvard University, through a grant (60477) from
the John Templeton Foundation; the China Scholarship
Council;  Consejo
Nacional de Ciencia y Tecnolog\'{\i}a (CONACYT,
Mexico, projects  U0004-246083, U0004-259839, F0003-272050, M0037-279006, F0003-281692,
104497, 275201, 263356);
the Delaney Family via the Delaney Family John A.
Wheeler Chair at Perimeter Institute; Dirección General
de Asuntos del Personal Académico-—Universidad
Nacional Autónoma de México (DGAPA-—UNAM,
projects IN112417 and IN112820); the European Research Council Synergy
Grant "BlackHoleCam: Imaging the Event Horizon
of Black Holes" (grant 610058); the Generalitat
Valenciana postdoctoral grant APOSTD/2018/177 and
GenT Program (project CIDEGENT/2018/021); MICINN Research Project PID2019-108995GB-C22;
the European Research Council for advanced grant 'JETSET: Launching, propagation and 
emission of relativistic 
jets from binary mergers and across mass scales' (Grant No. 884631); the
Gordon and Betty Moore Foundation (grant GBMF-3561); the Istituto Nazionale di Fisica
Nucleare (INFN) sezione di Napoli, iniziative specifiche
TEONGRAV; the two Dutch National Supercomputers, Cartesius and Snellius  
(NWO Grant 2021.013);
the International Max Planck Research
School for Astronomy and Astrophysics at the
Universities of Bonn and Cologne; 
Joint Princeton/Flatiron and Joint Columbia/Flatiron Postdoctoral Fellowships, 
research at the Flatiron Institute is supported by the Simons Foundation; 
the Japanese Government (Monbukagakusho:
MEXT) Scholarship; the Japan Society for
the Promotion of Science (JSPS) Grant-in-Aid for JSPS
Research Fellowship (JP17J08829); the Key Research
Program of Frontier Sciences, Chinese Academy of
Sciences (CAS, grants QYZDJ-SSW-SLH057, QYZDJSSW-
SYS008, ZDBS-LY-SLH011); the Leverhulme Trust Early Career Research
Fellowship; the Max-Planck-Gesellschaft (MPG);
the Max Planck Partner Group of the MPG and the
CAS; the MEXT/JSPS KAKENHI (grants 18KK0090, JP21H01137,
JP18H03721, 18K03709,
18H01245, 25120007); the Malaysian Fundamental Research Grant Scheme (FRGS) FRGS/1/2019/STG02/UM/02/6; the MIT International Science
and Technology Initiatives (MISTI) Funds; 
the Ministry of Science and Technology (MOST) of Taiwan (103-2119-M-001-010-MY2, 105-2112-M-001-025-MY3, 105-2119-M-001-042, 106-2112-M-001-011, 106-2119-M-001-013, 106-2119-M-001-027, 106-2923-M-001-005, 107-2119-M-001-017, 107-2119-M-001-020, 107-2119-M-001-041, 107-2119-M-110-005, 107-2923-M-001-009, 108-2112-M-001-048, 108-2112-M-001-051, 108-2923-M-001-002, 109-2112-M-001-025, 109-2124-M-001-005, 109-2923-M-001-001, 110-2112-M-003-007-MY2, 110-2112-M-001-033, 110-2124-M-001-007, and 110-2923-M-001-001);
the Ministry of Education (MoE) of Taiwan Yushan Young Scholar Program;
the Physics Division, National Center for Theoretical Sciences of Taiwan;
the National Aeronautics and
Space Administration (NASA, Fermi Guest Investigator
grant 80NSSC20K1567, NASA Astrophysics Theory Program grant 80NSSC20K0527, NASA NuSTAR award 80NSSC20K0645); 
the National
Institute of Natural Sciences (NINS) of Japan; the National
Key Research and Development Program of China
(grant 2016YFA0400704, 2016YFA0400702); the National
Science Foundation (NSF, grants AST-0096454,
AST-0352953, AST-0521233, AST-0705062, AST-0905844, AST-0922984, AST-1126433, AST-1140030,
DGE-1144085, AST-1207704, AST-1207730, AST-1207752, MRI-1228509, OPP-1248097, AST-1310896,  
AST-1555365, AST-1615796, AST-1715061, AST-1716327,  AST-1903847,AST-2034306); the Natural Science
Foundation of China (grants 
11650110427, 10625314, 11721303, 11725312, 11933007, 11991052, 11991053);
NWO grant number OCENW.KLEIN.113; a 
fellowship of China Postdoctoral Science Foundation (2020M671266); the Natural
Sciences and Engineering Research Council of
Canada (NSERC, including a Discovery Grant and
the NSERC Alexander Graham Bell Canada Graduate
Scholarships-Doctoral Program); the National Youth
Thousand Talents Program of China; the National Research
Foundation of Korea (the Global PhD Fellowship
Grant: grants NRF-2015H1A2A1033752, 2015-
R1D1A1A01056807, the Korea Research Fellowship Program:
NRF-2015H1D3A1066561, Basic Research Support Grant 2019R1F1A1059721); the Netherlands Organization
for Scientific Research (NWO) VICI award
(grant 639.043.513) and Spinoza Prize SPI 78-409; the
New Scientific Frontiers with Precision Radio Interferometry
Fellowship awarded by the South African Radio
Astronomy Observatory (SARAO), which is a facility
of the National Research Foundation (NRF), an
agency of the Department of Science and Technology
(DST) of South Africa; the Onsala Space Observatory
(OSO) national infrastructure, for the provisioning
of its facilities/observational support (OSO receives
funding through the Swedish Research Council under
grant 2017-00648) the Perimeter Institute for Theoretical
Physics (research at Perimeter Institute is supported
by the Government of Canada through the Department
of Innovation, Science and Economic Development
and by the Province of Ontario through the
Ministry of Research, Innovation and Science); 
the Spanish Ministerio de Ciencia e Innovación (grants PGC2018-098915-B-C21, AYA2016-80889-P,
PID2019-108995GB-C21, PID2020-117404GB-C21); 
the State
Agency for Research of the Spanish MCIU through
the "Center of Excellence Severo Ochoa" award for
the Instituto de Astrofísica de Andalucía (SEV-2017-
0709); the Toray Science Foundation; the Consejería de Economía, Conocimiento, 
Empresas y Universidad 
of the Junta de Andalucía (grant P18-FR-1769), the Consejo Superior de Investigaciones 
Científicas (grant 2019AEP112);
the M2FINDERS project which has received funding by the European Research Council (ERC) under 
the European Union’s Horizon 2020 Research and Innovation Programme (grant agreement No 101018682);
the US Department
of Energy (USDOE) through the Los Alamos National
Laboratory (operated by Triad National Security,
LLC, for the National Nuclear Security Administration
of the USDOE (Contract 89233218CNA000001);
 the European Union’s Horizon 2020
research and innovation programme under grant agreement
No 730562 RadioNet;
Shanghai Pilot Program for Basic Research, Chinese Academy of Science, 
Shanghai Branch (JCYJ-SHFY-2021-013);
ALMA North America Development
Fund; the Academia Sinica; Chandra DD7-18089X and TM6-
17006X; the GenT Program (Generalitat Valenciana)
Project CIDEGENT/2018/021. This work used the
Extreme Science and Engineering Discovery Environment
(XSEDE), supported by NSF grant ACI-1548562,
and CyVerse, supported by NSF grants DBI-0735191,
DBI-1265383, and DBI-1743442. XSEDE Stampede2 resource
at TACC was allocated through TG-AST170024
and TG-AST080026N. XSEDE JetStream resource at
PTI and TACC was allocated through AST170028.
The simulations were performed in part on the SuperMUC
cluster at the LRZ in Garching, on the
LOEWE cluster in CSC in Frankfurt, and on the
HazelHen cluster at the HLRS in Stuttgart. This
research was enabled in part by support provided
by Compute Ontario (http://computeontario.ca), Calcul
Quebec (http://www.calculquebec.ca) and Compute
Canada (http://www.computecanada.ca). 
CC acknowledges support from the Swedish Research Council (VR).
We thank
the staff at the participating observatories, correlation
centers, and institutions for their enthusiastic support.
This paper makes use of the following ALMA data:
ADS/JAO.ALMA\#2016.1.01154.V. ALMA is a partnership
of the European Southern Observatory (ESO;
Europe, representing its member states), NSF, and
National Institutes of Natural Sciences of Japan, together
with National Research Council (Canada), Ministry
of Science and Technology (MOST; Taiwan),
Academia Sinica Institute of Astronomy and Astrophysics
(ASIAA; Taiwan), and Korea Astronomy and
Space Science Institute (KASI; Republic of Korea), in
cooperation with the Republic of Chile. The Joint
ALMA Observatory is operated by ESO, Associated
Universities, Inc. (AUI)/NRAO, and the National Astronomical
Observatory of Japan (NAOJ). The NRAO
is a facility of the NSF operated under cooperative agreement
by AUI.
Hector Olivares and  Gibwa Musoke
were supported by Virtual Institute of Accretion (VIA) postdoctoral fellowships 
from the Netherlands Research School for Astronomy (NOVA).
APEX is a collaboration between the
Max-Planck-Institut f{\"u}r Radioastronomie (Germany),
ESO, and the Onsala Space Observatory (Sweden). The
SMA is a joint project between the SAO and ASIAA
and is funded by the Smithsonian Institution and the
Academia Sinica. The JCMT is operated by the East
Asian Observatory on behalf of the NAOJ, ASIAA, and
KASI, as well as the Ministry of Finance of China, Chinese
Academy of Sciences, and the National Key R\&D
Program (No. 2017YFA0402700) of China. Additional
funding support for the JCMT is provided by the Science
and Technologies Facility Council (UK) and participating
universities in the UK and Canada. 
Simulations were performed in part on the SuperMUC cluster at the LRZ in Garching, on the 
LOEWE cluster in CSC in Frankfurt, on the HazelHen cluster at the HLRS in Stuttgart, 
and on the Pi2.0 and Siyuan Mark-I at Shanghai Jiao Tong University.
The computer resources of the Finnish IT Center for Science (CSC) and the Finnish Computing 
Competence Infrastructure (FCCI) project are acknowledged.
Junghwan Oh was supported by Basic Science Research Program through the National Research
Foundation of Korea(NRF) funded by the Ministry of Education(NRF-2021R1A6A3A01086420).
We thank Martin Shepherd for the addition of extra features in the Difmap software 
that were used for the CLEAN imaging results presented in this paper.
The computing cluster of Shanghai VLBI correlator supported by the Special Fund 
for Astronomy from the Ministry of Finance in China is acknowledged.
The LMT is a project operated by the Instituto Nacional
de Astrófisica, Óptica, y Electrónica (Mexico) and the
University of Massachusetts at Amherst (USA). The
IRAM 30-m telescope on Pico Veleta, Spain is operated
by IRAM and supported by CNRS (Centre National de
la Recherche Scientifique, France), MPG (Max-Planck-
Gesellschaft, Germany) and IGN (Instituto Geográfico
Nacional, Spain). The SMT is operated by the Arizona
Radio Observatory, a part of the Steward Observatory
of the University of Arizona, with financial support of
operations from the State of Arizona and financial support
for instrumentation development from the NSF.
Support for SPT participation in the EHT is provided by 
the National Science Foundation through award OPP-1852617 
to the University of Chicago. Partial support is also 
provided by the Kavli Institute of Cosmological Physics 
at the University of Chicago. The SPT hydrogen maser was 
provided on loan from the GLT, courtesy of ASIAA.
Support for this work was provided by NASA through the NASA Hubble Fellowship grant
\#HST--HF2--51494.001 awarded by the Space Telescope Science Institute, which is operated 
by the Association of Universities for Research in Astronomy, Inc., for NASA, 
under contract NAS5--26555.
The EHTC has
received generous donations of FPGA chips from Xilinx
Inc., under the Xilinx University Program. The EHTC
has benefited from technology shared under open-source
license by the Collaboration for Astronomy Signal Processing
and Electronics Research (CASPER). The EHT
project is grateful to T4Science and Microsemi for their
assistance with Hydrogen Masers. This research has
made use of NASA’s Astrophysics Data System. We
gratefully acknowledge the support provided by the extended
staff of the ALMA, both from the inception of
the ALMA Phasing Project through the observational
campaigns of 2017 and 2018. We would like to thank
A. Deller and W. Brisken for EHT-specific support with
the use of DiFX. We thank J. Delgado for helpful discussions and feedback. We acknowledge the significance that
Maunakea, where the SMA and JCMT EHT stations
are located, has for the indigenous Hawaiian people.}

\bibliography{ref}

\begin{thebibliography}{49}
\expandafter\ifx\csname natexlab\endcsname\relax\def\natexlab#1{#1}\fi

\bibitem[{Akiyama {et~al.}(2017)Akiyama, Kuramochi, Ikeda, Fish, Tazaki, Honma,
  Doeleman, {E. Broderick}, Dexter, Mo{\'{s}}cibrodzka, Bouman, {A. Chael}, \&
  Zaizen}]{Akiyama_2017a}
Akiyama, K., {et~al.} 2017, ApJ, 838, 1

\bibitem[{Barber {et~al.}(1996)Barber, Dobkin, \& Huhdanpaa}]{Barber1996}
Barber, C.~B., Dobkin, D.~P., \& Huhdanpaa, H.~T. 1996, ACM Trans. on
  Mathematical Software, 22, 469

\bibitem[{Bouman {et~al.}(2018)Bouman, Johnson, Zoran, Fish, Doeleman, \&
  Freeman}]{Bouman2018}
Bouman, K.~L., Johnson, M.~D., Zoran, D., Fish, V.~L., Doeleman, S.~S., \&
  Freeman, W.~T. 2018, in Proceedings of the IEEE Computer Society Conference
  on Computer Vision and Pattern Recognition

\bibitem[{Chael {et~al.}(2018)Chael, Johnson, Bouman, Blackburn, Akiyama, \&
  Narayan}]{Chael2018}
Chael, A.~A., Johnson, M.~D., Bouman, K.~L., Blackburn, L.~L., Akiyama, K., \&
  Narayan, R. 2018, ApJ, 857, 23

\bibitem[{Chael {et~al.}(2016)Chael, Johnson, Narayan, Doeleman, Wardle, \&
  Bouman}]{Chael_2016}
Chael, A.~A., Johnson, M.~D., Narayan, R., Doeleman, S.~S., Wardle, J. F.~C.,
  \& Bouman, K.~L. 2016, ApJ, 829, 11

\bibitem[{Clark(1980)}]{Clark1980}
Clark, B. 1980, A\&A, 89, 377

\bibitem[{Cornwell \& Wilkinson(1981)}]{Cornwell1981}
Cornwell, T.~J., \& Wilkinson, P.~N. 1981, MNRAS, 196, 1067

\bibitem[{Do {et~al.}(2019)Do, Hees, Ghez, Martinez, Chu, Jia, Sakai, Lu,
  Gautam, O'Neil, Becklin, Morris, Matthews, Nishiyama, Campbell, Chappell,
  Chen, Ciurlo, Dehghanfar, Gallego-Cano, Kerzendorf, Lyke, Naoz, Saida,
  Sch{\"{o}}del, Takahashi, Takamori, Witzel, \& Wizinowich}]{Do2019}
Do, T., {et~al.} 2019, Science, 365, 664

\bibitem[{Doeleman {et~al.}(2008)Doeleman, Weintroub, Rogers, Plambeck, Freund,
  Tilanus, Friberg, Ziurys, Moran, Corey, Young, Smythe, Titus, Marrone,
  Cappallo, Bock, Bower, Chamberlin, Davis, Krichbaum, Lamb, Maness, Niell,
  Roy, Strittmatter, Werthimer, Whitney, \& Woody}]{Doeleman2008}
Doeleman, S.~S., {et~al.} 2008, Nature, 455

\bibitem[{{Event Horizon Telescope Collaboration}
  {et~al.}(2019{\natexlab{a}}){Event Horizon Telescope Collaboration}, Akiyama,
  Alberdi, Alef, Asada, Azulay, Baczko, Ball, Balokovi{\'{c}}, Barrett,
  Bintley, Blackburn, Boland, Bouman, Bower, Bremer, Brinkerink, Brissenden,
  Britzen, Broderick, Broguiere, Bronzwaer, Byun, Carlstrom, Chael, \&
  al.}]{EHT1}
{Event Horizon Telescope Collaboration} {et~al.} 2019{\natexlab{a}}, 875, L1

\bibitem[{{Event Horizon Telescope Collaboration}
  {et~al.}(2019{\natexlab{b}}){Event Horizon Telescope Collaboration}, Akiyama,
  Alberdi, Alef, Asada, Azulay, Baczko, Ball, Balokovi{\'{c}}, Barrett,
  Bintley, Blackburn, Boland, Bouman, Bower, Bremer, Brinkerink, Brissenden,
  Britzen, Broderick, Broguiere, Bronzwaer, Byun, Carlstrom, Chael, \&
  al.}]{EHT2}
---. 2019{\natexlab{b}}, 875, L2

\bibitem[{{Event Horizon Telescope Collaboration}
  {et~al.}(2019{\natexlab{c}}){Event Horizon Telescope Collaboration}, Akiyama,
  Alberdi, Alef, Asada, Azulay, Baczko, Ball, Balokovi{\'{c}}, Barrett,
  Bintley, Blackburn, Boland, Bouman, Bower, Bremer, Brinkerink, Brissenden,
  Britzen, Broderick, Broguiere, Bronzwaer, Byun, Carlstrom, Chael, \&
  al.}]{EHT3}
---. 2019{\natexlab{c}}, 875, L3

\bibitem[{{Event Horizon Telescope Collaboration}
  {et~al.}(2019{\natexlab{d}}){Event Horizon Telescope Collaboration}, Akiyama,
  Alberdi, Alef, Asada, Azulay, Baczko, Ball, Balokovi{\'{c}}, Barrett,
  Bintley, Blackburn, Boland, Bouman, Bower, Bremer, Brinkerink, Brissenden,
  Britzen, Broderick, Broguiere, Bronzwaer, Byun, Carlstrom, Chael, \&
  al.}]{EHT4}
---. 2019{\natexlab{d}}, 875, L4

\bibitem[{{Event Horizon Telescope Collaboration}
  {et~al.}(2019{\natexlab{e}}){Event Horizon Telescope Collaboration}, Akiyama,
  Alberdi, Alef, Asada, Azulay, Baczko, Ball, Balokovi{\'{c}}, Barrett,
  Bintley, Blackburn, Boland, Bouman, Bower, Bremer, Brinkerink, Brissenden,
  Britzen, Broderick, Broguiere, Bronzwaer, Byun, Carlstrom, Chael, \&
  al.}]{EHT5}
---. 2019{\natexlab{e}}, 875, L5

\bibitem[{{Event Horizon Telescope Collaboration}
  {et~al.}(2019{\natexlab{f}}){Event Horizon Telescope Collaboration}, Akiyama,
  Alberdi, Alef, Asada, Azulay, Baczko, Ball, Balokovi{\'{c}}, Barrett,
  Bintley, Blackburn, Boland, Bouman, Bower, Bremer, Brinkerink, Brissenden,
  Britzen, Broderick, Broguiere, Bronzwaer, Byun, Carlstrom, Chael, \&
  al.}]{EHT6}
---. 2019{\natexlab{f}}, 875, L6

\bibitem[{{Event Horizon Telescope Collaboration}
  {et~al.}(2022{\natexlab{a}}){Event Horizon Telescope Collaboration},
  {Akiyama}, {Alberdi}, {Alef}, {Asada}, {Azulay}, {Baczko}, {Ball},
  {Balokovi{\'c}}, {Barrett}, \& et~al.}]{PaperI}
---. 2022{\natexlab{a}}, submitted to \apjl, (Paper I)

\bibitem[{{Event Horizon Telescope Collaboration}
  {et~al.}(2022{\natexlab{b}}){Event Horizon Telescope Collaboration},
  {Akiyama}, {Alberdi}, {Alef}, {Asada}, {Azulay}, {Baczko}, {Ball},
  {Balokovi{\'c}}, {Barrett}, \& et~al.}]{PaperII}
---. 2022{\natexlab{b}}, submitted to \apjl, (Paper II)

\bibitem[{{Event Horizon Telescope Collaboration}
  {et~al.}(2022{\natexlab{c}}){Event Horizon Telescope Collaboration},
  {Akiyama}, {Alberdi}, {Alef}, {Asada}, {Azulay}, {Baczko}, {Ball},
  {Balokovi{\'c}}, {Barrett}, \& et~al.}]{PaperIII}
---. 2022{\natexlab{c}}, submitted to \apjl, (Paper III)

\bibitem[{{Event Horizon Telescope Collaboration}
  {et~al.}(2022{\natexlab{d}}){Event Horizon Telescope Collaboration},
  {Akiyama}, {Alberdi}, {Alef}, {Asada}, {Azulay}, {Baczko}, {Ball},
  {Balokovi{\'c}}, {Barrett}, \& et~al.}]{PaperIV}
---. 2022{\natexlab{d}}, submitted to \apjl, (Paper IV)

\bibitem[{{Event Horizon Telescope Collaboration}
  {et~al.}(2022{\natexlab{e}}){Event Horizon Telescope Collaboration},
  {Akiyama}, {Alberdi}, {Alef}, {Asada}, {Azulay}, {Baczko}, {Ball},
  {Balokovi{\'c}}, {Barrett}, \& et~al.}]{PaperV}
---. 2022{\natexlab{e}}, submitted to \apjl, (Paper V)

\bibitem[{{Event Horizon Telescope Collaboration}
  {et~al.}(2022{\natexlab{f}}){Event Horizon Telescope Collaboration},
  {Akiyama}, {Alberdi}, {Alef}, {Asada}, {Azulay}, {Baczko}, {Ball},
  {Balokovi{\'c}}, {Barrett}, \& et~al.}]{PaperVI}
---. 2022{\natexlab{f}}, submitted to \apjl, (Paper VI)

\bibitem[{Fried(1978)}]{Fried1978}
Fried, D.~L. 1978, J Opt Soc Am, 68, 1651

\bibitem[{{GRAVITY Collaboration} {et~al.}(2018{\natexlab{a}}){GRAVITY
  Collaboration}, {Abuter, R.}, {Amorim, A.}, {Baub{\"{o}}ck, M.}, {Berger, J.
  P.}, {Bonnet, H.}, {Brandner, W.}, {Cl{\'{e}}net, Y.}, {Coud{\'{e}} du
  Foresto, V.}, {de Zeeuw, P. T.}, {Deen, C.}, {Dexter, J.}, {Duvert, G.},
  {Eckart, A.}, {Eisenhauer, F.}, {F{\"{o}}rster Schreiber, N. M.}, {Garcia,
  P.}, {Gao, F.}, {Gendron, E.}, {Genzel, R.}, {Gillessen, S.}, {Guajardo, P.},
  {Habibi, M.}, {Haubois, X.}, {Henning, Th.}, {Hippler, S.}, {Horrobin, M.},
  {Huber, A.}, {Jim{\'{e}}nez-Rosales, A.}, {Jocou, L.}, {Kervella, P.},
  {Lacour, S.}, {Lapeyr{\`{e}}re, V.}, {Lazareff, B.}, {Le Bouquin, J.-B.},
  {L{\'{e}}na, P.}, {Lippa, M.}, {Ott, T.}, {Panduro, J.}, {Paumard, T.},
  {Perraut, K.}, {Perrin, G.}, {Pfuhl, O.}, {Plewa, P. M.}, {Rabien, S.},
  {Rodr\'\ iguez-Coira, G.}, {Rousset, G.}, {Sternberg, A.}, {Straub, O.},
  {Straubmeier, C.}, {Sturm, E.}, {Tacconi, L. J.}, {Vincent, F.}, {von
  Fellenberg, S.}, {Waisberg, I.}, {Widmann, F.}, {Wieprecht, E.}, {Wiezorrek,
  E.}, {Woillez, J.}, \& {Yazici, S.}}]{gravity_2018_hotspot}
{GRAVITY Collaboration} {et~al.} 2018{\natexlab{a}}, A\&A, 618, L10

\bibitem[{{GRAVITY Collaboration} {et~al.}(2018{\natexlab{b}}){GRAVITY
  Collaboration}, {Abuter, R.}, {Amorim, A.}, {Anugu, N.}, {Baub{\"{o}}ck, M.},
  {Benisty, M.}, {Berger, J. P.}, {Blind, N.}, {Bonnet, H.}, {Brandner, W.},
  {Buron, A.}, {Collin, C.}, {Chapron, F.}, {Cl{\'{e}}net, Y.}, {dCoud{\'{e}} u
  Foresto, V.}, {de Zeeuw, P. T.}, {Deen, C.}, {Delplancke-Str{\"{o}}bele, F.},
  {Dembet, R.}, {Dexter, J.}, {Duvert, G.}, {Eckart, A.}, {Eisenhauer, F.},
  {Finger, G.}, {F{\"{o}}rster Schreiber, N. M.}, {F{\'{e}}dou, P.}, {Garcia,
  P.}, {Garcia Lopez, R.}, {Gao, F.}, {Gendron, E.}, {Genzel, R.}, {Gillessen,
  S.}, {Gordo, P.}, {Habibi, M.}, {Haubois, X.}, {Haug, M.}, {Hausmann, F.},
  {Henning, Th.}, {Hippler, S.}, {Horrobin, M.}, {Hubert, Z.}, {Hubin, N.},
  {Jimenez Rosales, A.}, {Jochum, L.}, {Jocou, L.}, {Kaufer, A.}, {Kellner,
  S.}, {Kendrew, S.}, {Kervella, P.}, {Kok, Y.}, {Kulas, M.}, {Lacour, S.},
  {Lapeyr{\`{e}}re, V.}, {Lazareff, B.}, {Le Bouquin, J.-B.}, {L{\'{e}}na, P.},
  {Lippa, M.}, {Lenzen, R.}, {M{\'{e}}rand, A.}, {M{\"{u}}ler, E.}, {Neumann,
  U.}, {Ott, T.}, {Palanca, L.}, {Paumard, T.}, {Pasquini, L.}, {Perraut, K.},
  {Perrin, G.}, {Pfuhl, O.}, {Plewa, P. M.}, {Rabien, S.}, {Ram\'\ irez, A.},
  {Ramos, J.}, {Rau, C.}, {Rodr\'\ iguez-Coira, G.}, {Rohloff, R.-R.},
  {Rousset, G.}, {Sanchez-Bermudez, J.}, {Scheithauer, S.}, {Sch{\"{o}}ller,
  M.}, {Schuler, N.}, {Spyromilio, J.}, {Straub, O.}, {Straubmeier, C.},
  {Sturm, E.}, {Tacconi, L. J.}, {Tristram, K. R. W.}, {Vincent, F.}, {von
  Fellenberg, S.}, {Wank, I.}, {Waisberg, I.}, {Widmann, F.}, {Wieprecht, E.},
  {Wiest, M.}, {Wiezorrek, E.}, {Woillez, J.}, {Yazici, S.}, {Ziegler, D.}, \&
  {Zins, G.}}]{gravity_2018_s2}
---. 2018{\natexlab{b}}, A\&A, 615, L15

\bibitem[{H{\"{o}}gbom(1974)}]{Hogbom_1974}
H{\"{o}}gbom, J.~A. 1974, A\&ASS, 15, 417

\bibitem[{Honma {et~al.}(2014)Honma, Akiyama, Uemura, \& Ikeda}]{Honma_2014}
Honma, M., Akiyama, K., Uemura, M., \& Ikeda, S. 2014, PASP, 66

\bibitem[{Issaoun {et~al.}(2019)Issaoun, Johnson, Blackburn, Brinkerink,
  Mo{\'{s}}cibrodzka, Chael, Goddi, Mart{\'{i}}-Vidal, Wagner, Doeleman,
  Falcke, Krichbaum, Akiyama, Bach, Bouman, Bower, Broderick, Cho, Crew,
  Dexter, Fish, Gold, G{\'{o}}mez, Hada, Hern{\'{a}}ndez-G{\'{o}}mez,
  Jan{\ss}en, Kino, Kramer, Loinard, Lu, Markoff, Marrone, Matthews, Moran,
  M{\"{u}}ller, Roelofs, Ros, Rottmann, Sanchez, Tilanus, de~Vicente, Wielgus,
  Zensus, \& Zhao}]{Issaoun2019}
Issaoun, S., {et~al.} 2019, ApJ, 871

\bibitem[{Johannsen \& Psaltis(2010)}]{Johannsen2010}
Johannsen, T., \& Psaltis, D. 2010, Astrophysical Journal, 718, 446

\bibitem[{Johnson {et~al.}(2017)Johnson, Bouman, Blackburn, Chael, Rosen,
  Shiokawa, Roelofs, Akiyama, Fish, \& Doeleman}]{Johnson2017}
Johnson, M.~D., {et~al.} 2017, ApJ, 850, 172

\bibitem[{Kamruddin \& Dexter(2013)}]{Kamruddin2013}
Kamruddin, A.~B., \& Dexter, J. 2013, Monthly Notices of the Royal Astronomical
  Society, 434, 765

\bibitem[{Kullback \& Leibler(1951)}]{kullback_1951}
Kullback, S., \& Leibler, R.~A. 1951, The Annals of Mathematical Statistics,
  22, 79

\bibitem[{Kuramochi {et~al.}(2018)Kuramochi, Akiyama, Ikeda, Tazaki, Fish, Pu,
  Asada, \& Honma}]{Kuramochi2018}
Kuramochi, K., Akiyama, K., Ikeda, S., Tazaki, F., Fish, V.~L., Pu, H.-Y.,
  Asada, K., \& Honma, M. 2018, ApJ, 858, 56

\bibitem[{Lal \& Lobanov(2007)}]{Lal_2010}
Lal, D.~V., \& Lobanov, A.~P. 2007, in Proceedings of Science, Vol.~52

\bibitem[{Mart{\'{i}}-Vidal {et~al.}(2011)Mart{\'{i}}-Vidal, Marcaide, Alberdi,
  P{\'{e}}rez-Torres, Ros, \& Guirado}]{Marti-Vidal2011}
Mart{\'{i}}-Vidal, I., Marcaide, J.~M., Alberdi, A., P{\'{e}}rez-Torres, M.~A.,
  Ros, E., \& Guirado, J.~C. 2011, Astronomy and Astrophysics, 533

\bibitem[{Massi {et~al.}(2012)Massi, Ros, \& Zimmermann}]{Massi2012}
Massi, M., Ros, E., \& Zimmermann, L. 2012, Astronomy and Astrophysics, 540

\bibitem[{Miller-Jones {et~al.}(2019)Miller-Jones, Tetarenko, Sivakoff,
  Middleton, Altamirano, Anderson, Belloni, Fender, Jonker, K{\"{o}}rding,
  Krimm, Maitra, Markoff, Migliari, Mooley, Rupen, Russell, Russell, Sarazin,
  Soria, \& Tudose}]{Miller2019}
Miller-Jones, J.~C., {et~al.} 2019, \nat, 569, 374

\bibitem[{Narayan \& Nityananda(1986)}]{Narayan1986}
Narayan, R., \& Nityananda, R. 1986, Annual Review of A\&A, 24, 127

\bibitem[{Palumbo {et~al.}(2019)Palumbo, Doeleman, Johnson, Bouman, \&
  Chael}]{Palumbo_2019}
Palumbo, D.~C., Doeleman, S.~S., Johnson, M.~D., Bouman, K.~L., \& Chael, A.~A.
  2019, \apj, 881, 62

\bibitem[{Pearson \& Readhead(1984)}]{Pearson1984}
Pearson, T.~J., \& Readhead, A. C.~S. 1984, Annual Review of A\&A, 22, 97

\bibitem[{Psaltis(2019)}]{Psaltis2018}
Psaltis, D. 2019, General Relativity and Gravitation, 51

\bibitem[{Psaltis {et~al.}(2015)Psaltis, {\"{O}}zel, Chan, \&
  Marrone}]{Psaltis2015}
Psaltis, D., {\"{O}}zel, F., Chan, C.~K., \& Marrone, D.~P. 2015, Astrophysical
  Journal, 814, 115

\bibitem[{Raymond {et~al.}(2021)Raymond, Palumbo, Paine, Blackburn,
  {C{\'{o}}rdova Rosado}, Doeleman, Farah, Johnson, Roelofs, Tilanus, \&
  Weintroub}]{Raymond_2021}
Raymond, A.~W., {et~al.} 2021, \apjs, 253, 5

\bibitem[{Readhead {et~al.}(1980)Readhead, Walker, Pearson, \&
  Cohen}]{Readhead1980}
Readhead, A.~C., Walker, R.~C., Pearson, T.~J., \& Cohen, M.~H. 1980, Nature,
  285, 137

\bibitem[{Shepherd(2011)}]{shepherd_2011}
Shepherd, M. 2011, {Difmap: Synthesis Imaging of Visibility Data}

\bibitem[{Shepherd(1997)}]{shepherd1997a}
Shepherd, M.~C. 1997, {Software and Systems VI}

\bibitem[{Thompson {et~al.}(2017)Thompson, Moran, \& {Swenson George W.}}]{TMS}
Thompson, A.~R., Moran, J.~M., \& {Swenson George W.}, J. 2017, {Interferometry
  and Synthesis in Radio Astronomy, 3rd Edition}

\bibitem[{Wielgus {et~al.}(2020{\natexlab{a}})Wielgus, Palumbo, \&
  Hamilton}]{github_LCG}
Wielgus, M., Palumbo, D., \& Hamilton, L. 2020{\natexlab{a}}, {LCG metric},
  \url{https://github.com/wielgusm/mwtools/blob/master/LCG_metric.py}

\bibitem[{Wielgus {et~al.}(2020{\natexlab{b}})Wielgus, Akiyama, Blackburn,
  Chan, Dexter, Doeleman, Fish, Issaoun, \& Johnson}]{Wielgus2019}
Wielgus, M., {et~al.} 2020{\natexlab{b}}, \apj, 901, 67

\bibitem[{Wilkinson {et~al.}(1977)Wilkinson, Readhead, Purcell, \&
  Anderson}]{Wilkinson1977}
Wilkinson, P.~N., Readhead, A.~C., Purcell, G.~H., \& Anderson, B. 1977,
  Nature, 269, 764

\end{thebibliography}

\clearpage
\appendix
\section{Sample synthetic data products}
\label{sec:sample_data}
Here, we present representative samples of the visibility amplitudes and closure phases that are generated from the synthetic data described in \autoref{sec:models}.

\begin{figure}[h]
\centering
\includegraphics[width=\columnwidth]{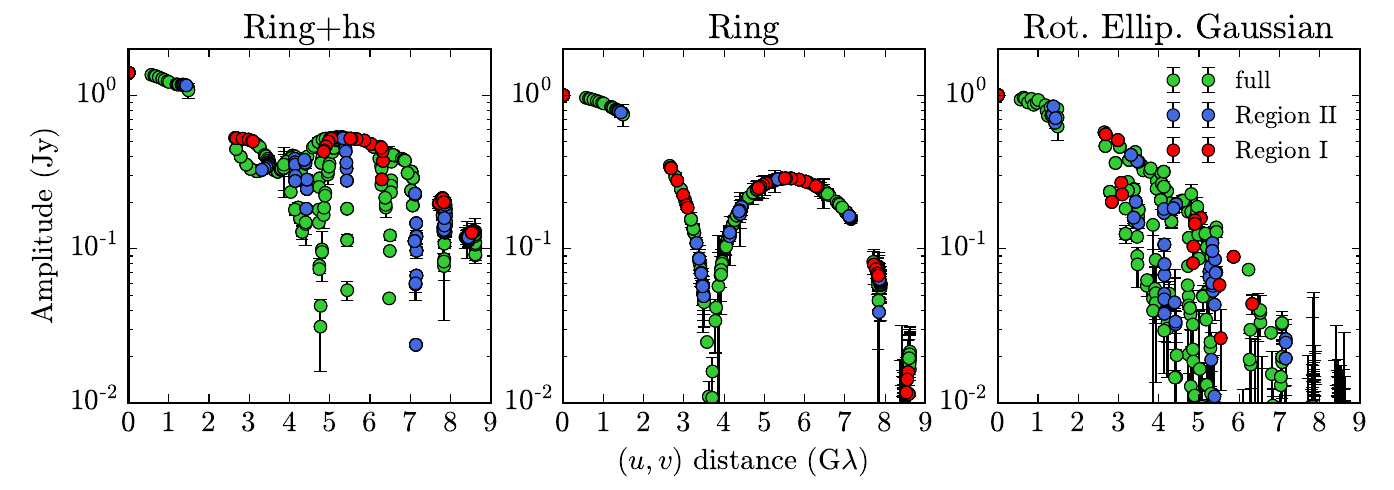}
\caption{Example amplitudes for the three main synthetic data types described in \autoref{sec:models}. The time-variable models have periods of 270 minutes. The amplitudes recorded during Region I and Region II (see \autoref{tab:timestamps}) are shown in red and blue, respectively. Region II has higher radial homogeneity in \uv distance, which contributes to its higher metric score and increased dynamical imaging capability. Errorbars show $1\sigma$ thermal noise.} 
\label{fig:appendix_amps}
\end{figure}

\begin{figure}[h]
\centering
\includegraphics[width=\columnwidth]{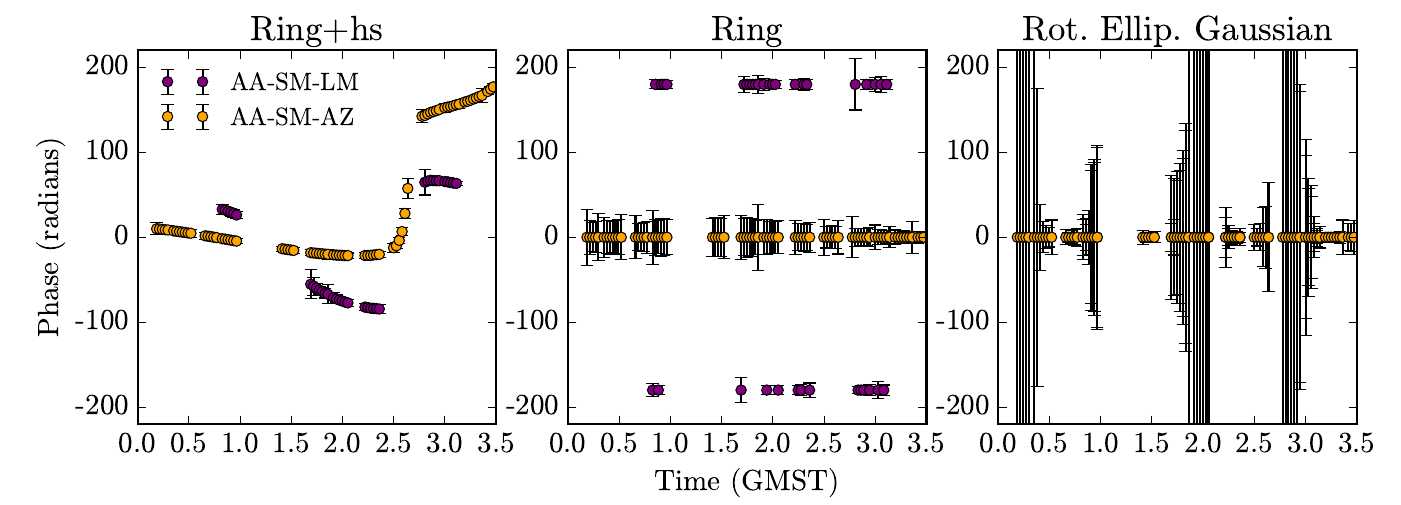}
\caption{Example closure phases on two triangles for the three main synthetic data types described in \autoref{sec:models}. The time-variable models have periods of 270 minutes. Closure phases are useful for identifying and constraining asymmetry and time variability in the source.} 
\label{fig:appendix_cphases}
\end{figure}

\section{Derivation of isotropy-based coverage metric}
\label{sec:derivation}

Section \ref{sec:demo} demonstrated that quantifying the isotropy of a \uv coverage configuration can indicate whether it is suitable for producing accurate reconstructions of a dynamic source. We adopt a coverage metric of the form 
\begin{align}
    \mathcal{C} = I(\{\vec{u}_i\})R(\{\vec{u}_i\}),
    \label{eq:coverage_parameter_general}
\end{align}
where $\{\vec{u}_i\}=\{(u_i, v_i)\}$ is the set of $2N$ baselines (including their Hermitian conjugates), $I$ is a measure of the isotropy of the coverage, and $R$ is a measure of the radial homogeneity of the coverage. 

To estimate the radial homogeneity, we compare the cumulative distribution function (CDF) of the distribution of baseline lengths against the uniform CDF via the Kolmogorov-Smirnov (KS) test. The uniform distribution examined for this test ranges from 0 G$\lambda$ to the maximum baseline length achieved in the observation. This test returns a ``distance'' $\mathcal{K}$ between the distributions, which increases as the distribution becomes less radially homogeneous, making use of the test in this context a measure of radial inhomogeneity. To convert the result of the KS test into a measure of radial homogeneity, we select an upper bound $\mathcal{K}_{\textrm{max}}$ corresponding to the maximum distance from uniform any individual baseline distribution obtains throughout the observation. We then subtract the result of the KS test from this maximum, i.e.,
\begin{align}
    \mathcal{K}' = \frac{\mathcal{K}_{\textrm{max}} - \mathcal{K}}{\mathcal{K}_{\textrm{max}}} = 1-\frac{\mathcal{K}}{\mathcal{K}_{\textrm{max}}},
    \label{eq:radial_homogeneity}
\end{align}
where $R(\{\vec{u}_i\})=\mathcal{K}'$ is our new metric of homogeneity. This metric is conveniently bounded between 0 and 1.  To make this metric absolute, a fixed value of $\mathcal{K}_{\textrm{max}}$ can be chosen arbitrarily and applied to multiple observations. For the absolute comparisons in this paper (see e.g., \autoref{sec:inter-day}) a value of  $\mathcal{K}_{\textrm{max}}=0.513338437261774\approx0.513$ is adopted. This value of $\mathcal{K}_{\textrm{max}}$ is chosen to be the maximum value of $\mathcal{K}$ achieved during the April 7 observation.

In order to measure the isotropy of the coverage, we examine the second moment (moment of inertia) of the distribution of baselines. As a spatial configuration of points with uniform weighting, the \uv coverage can be treated as a mass distribution. For a two-dimensional mass distribution, a disk is considered isotropic, and a rod is considered anisotropic, and the spectrum between the two cases can be probed using the moment of inertia tensor. Given $2N$ baselines and conjugate baselines with coordinates $\{\vec{u}_i\}=\{(u_i, v_i)\}$, we can compute the second moments of the distribution as
\begin{align}
    \langle u^2 \rangle &= \frac{1}{2N}\sum_i^{2N}u_i^2, \nonumber \\
    \langle v^2 \rangle &= \frac{1}{2N}\sum_i^{2N}v_i^2, \nonumber \\
    \langle uv \rangle &= \frac{1}{2N}\sum_i^{2N}u_iv_i.
\end{align}
The moment of inertia tensor $\mathcal{M}$ is
\begin{align}
    \mathcal{M} = \begin{bmatrix}
    \langle u^2 \rangle & \langle uv \rangle \\
    \langle uv \rangle & \langle v^2 \rangle
    \end{bmatrix}.
\end{align}
The principal moments of inertia can be computed from the eigenvalues $\lambda_1$ and $\lambda_2$ of $\mathcal{M}$. From these, we can derive the following orientation-independent measure of isotropy
\begin{align}
    \tilde{I}(\{\vec{u}_i\})=1-\frac{|\lambda_1-\lambda_2|}{\lambda_1+\lambda_2} =1-\frac{\sqrt{(\langle u^2 \rangle-\langle v^2 \rangle)^2+4\langle uv \rangle^2}}{\langle u^2 \rangle+\langle v^2 \rangle}.
\end{align}
Note that $\tilde{I}(\{\vec{u}_i\})$ can also be written in terms of the FWHM of the dirty beam,
\begin{align}
    \tilde{I}(\{\vec{u}_i\}) &= 1-\frac{1/\theta_{\rm min}^2-1/\theta_{\rm maj}^2}{1/\theta_{\rm min}^2+1/\theta_{\rm maj}^2}\\
    &= \frac{2 \theta_{\rm min}^2}{\theta_{\rm maj}^2 + \theta_{\rm min}^2}.
\end{align}

This measure of isotropy is naturally normalized between 0 and 1. Substituting this expression and \autoref{eq:radial_homogeneity} into \autoref{eq:coverage_parameter_general} gives the following expression for a coverage quality metric:
\begin{align}
    \mathcal{C} &= I(\{\vec{u}_i\})R(\{\vec{u}_i\}) \nonumber \\
    &= \left(1-\frac{\sqrt{(\langle u^2 \rangle-\langle v^2 \rangle)^2+4\langle uv \rangle^2}}{\langle u^2 \rangle+\langle v^2 \rangle}\right)\left(1-\frac{\mathcal{K}}{\mathcal{K}_{\textrm{max}}}\right)
\end{align}
This measure of coverage quality can be applied to partition an arbitrary VLBI observation into time regions ranked by their ability to accurately reconstruct dynamical sources.

\section{Sampled results from synthetic data reconstructions}
\label{sec:appendix_sampled_results}
\begin{figure*}[h]
\centering
\includegraphics[width=0.68\textwidth]{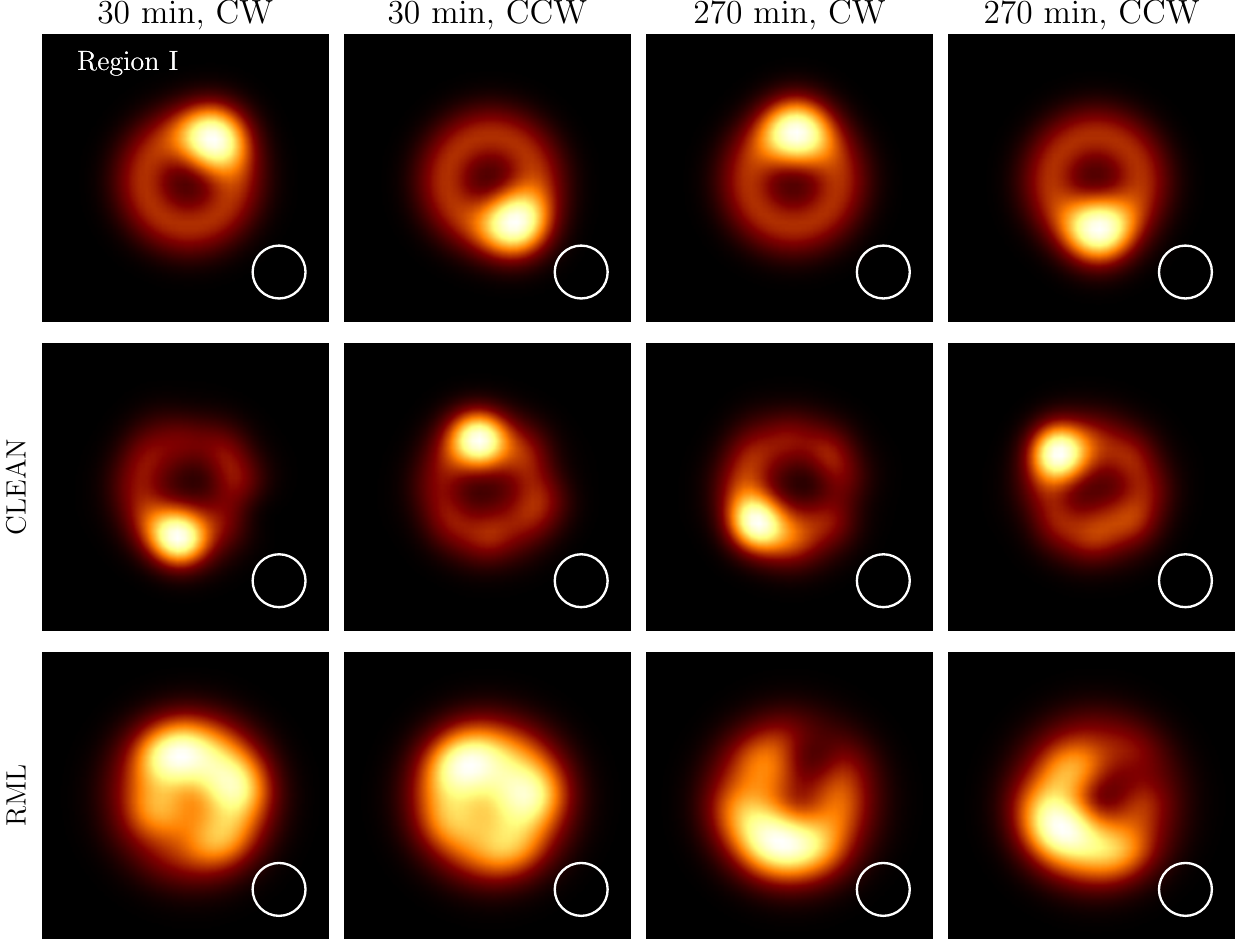}
\caption{Sampled reconstructions from Region I. The top row shows the true model image for each configuration. All panels show model images and reconstructions at ${\sim}$21:00 GMST on April 7. The white circle in the bottom right of each panel corresponds to an 18 $\mu$as diameter CLEAN beam. Even with substantial prior assumptions that facilitate ring reconstruction, the hotspot is frequently placed incorrectly, rendering this time region unsuitable for recovery of orbital angular variability.} 
\label{fig:regionI_sampled_reco}
\end{figure*}

\begin{figure*}[h]
\centering
\includegraphics[width=0.68\textwidth]{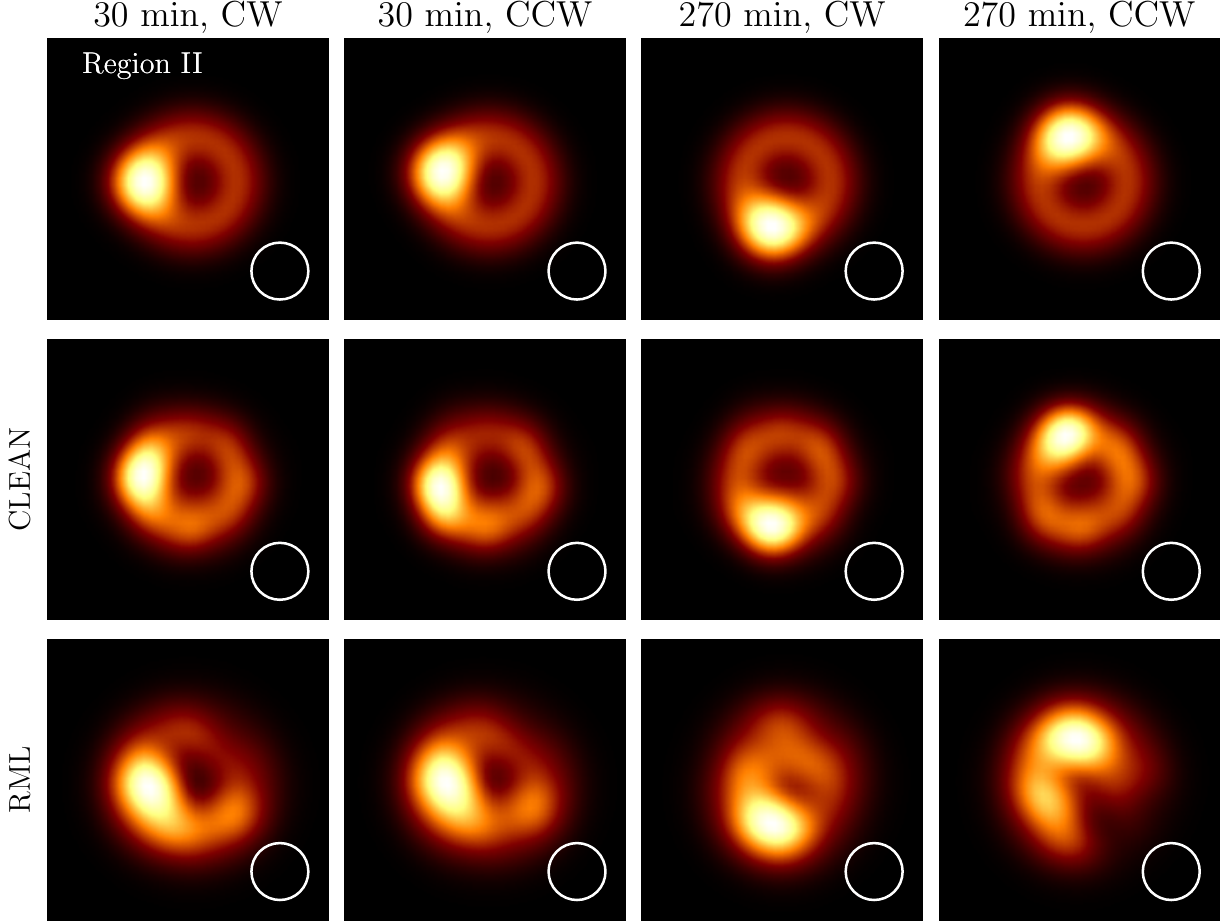}
\caption{Sampled reconstructions from Region II. The top row shows the true model image for each configuration. The white circle in the bottom right of each panel corresponds to an 18 $\mu$as diameter CLEAN beam. All panels show model images and reconstructions and reconstructions at ${\sim}$1:50 GMST on April 8. 
} 
\label{fig:regionII_sampled_reco}
\end{figure*}

In \autoref{sec:application_2017}, we test the coverage metrics examined in \autoref{sec:alternatives} by performing reconstructions of synthetic observations in selected time regions of the 2017 EHT observational coverage of \sgra. Here, we provide representative snapshot images from each of the time regions. Sampled snapshot images from Region I are shown in \autoref{fig:regionI_sampled_reco} and sampled snapshot images from Region II are shown in \autoref{fig:regionII_sampled_reco}. 

Reconstructions in both Region I and Region II demonstrate clear recovery of a ring-like feature of approximately 50 $\mu$as in diameter, as both sufficiently probe the radial distribution of the source Fourier transform to constrain the overall size of the source. However, the directionally-biased coverage of Region I produces incorrect reconstructions of hotspot location. By contrast, reconstructions in Region II repeatedly recovery the correct hotspot location across all periods, directions, and imaging algorithms. 

Though both time regions recover a ring-like feature of the approximately correct size, the Region II reconstructions provide a more accurate ring-to-central-depression flux density ratio. The increased accuracy is present in both the CLEAN and RML reconstructions. By contrast, reconstructions in Region I fail to consistently provide a visually distinctive depression and misrepresent the angular brightness profile of the source. 

\end{document}